\documentclass[aip,rsi,reprint,graphicx]{revtex4-1} 

\usepackage{amsmath}
\usepackage{amssymb}
\usepackage{graphicx} 
\usepackage{dcolumn}
\usepackage{bm} 
\usepackage{color}
\usepackage[colorlinks=true,linkcolor=blue]{hyperref}
\usepackage{epstopdf}

\draft 

\begin{document}


\title{Self-consistent iteration procedure in analyzing reflectivity and spectroscopic ellipsometry data of multilayered materials and their interfaces}

\author{T. C. Asmara}%
\affiliation{NUSNNI-NanoCore, Singapore Synchrotron Light Source, and Department of Physics, National University of Singapore, Singapore 117576}

\author{I. Santoso}%
\affiliation{NUSNNI-NanoCore, Singapore Synchrotron Light Source, and Department of Physics, National University of Singapore, Singapore 117576}

\author{A. Rusydi}%
\email[Correspondence to: ]{phyandri@nus.edu.sg}
\affiliation{NUSNNI-NanoCore, Singapore Synchrotron Light Source, and Department of Physics, National University of Singapore, Singapore 117576}

\date{\today}

\begin{abstract}
For multilayered materials, reflectivity depends on the complex dielectric function of all the constituent layers, and a detailed analysis is required to separate them. Furthermore, for some cases, new quantum states can occur at the interface which may change the optical properties of the material. In this paper, we discuss various aspects of such analysis, and present a self-consistent iteration procedure, a versatile method to extract and separate the complex dielectric function of each individual layer of a multilayered system. As a case study, we apply this method to LaAlO$_{3}$/SrTiO$_{3}$ heterostructure in which we are able to separate the effects of the interface from the LaAlO$_{3}$ film and the SrTiO$_{3}$ substrate. Our method can be applied to other complex multilayered systems with various numbers of layers.
\end{abstract}

\pacs{78.20.-e, 07.05.Kf, 73.21.Ac}

\maketitle 

\section{Introduction}

Recent technological advances in synthesizing multilayered materials with precise atomic control have made it possible to study the various exotic quantum phenomena that can occur at the interfaces of dissimilar materials \cite{HwangNatMat2012}. The interplay between charge, spin, and orbital at these interfaces can lead to various exciting phenomena such as orbital and spin reconstructions, metal-insulator transitions, magneto-electric coupling, superconductivity, quantum Hall effect, and topological effects. What makes it even more interesting is that these phenomena can occur even if the parent materials that make up the interface are not known to exhibit those properties. Examples include the ferromagnetic interface (and associated colossal magnetoresistance) between antiferromagnetic insulators LaMnO$_{3}$ and SrMnO$_{3}$ \cite{SalvadorAPL1999,YamadaAPL2006}, superconducting interface between insulator La$_{2}$CuO$_{4}$ and metallic overdoped (La,Sr)$_{2}$CuO$_{4}$ \cite{GozarNature2008}, conducting interface between band insulators LaAlO$_{3}$ (LAO) and SrTiO$_{3}$ (STO) \cite{OhtomoNature2004}, and many others \cite{HarrisonPRB1978,TakahashiAPL2001,OhtomoNature2002,YamadaAPL2002,LeeNature2005,ChakhalianNatPhys2006,TsukazakiScience2007,ChakhalianScience2007,KozukaNature2009,HiguchiPRB2009,YuPRL2010,BenckiserNatMater2011}. Another case is graphene; since it usually needs to be suspended on top of a substrate due to its two-dimensionality, it is also effectively a multilayered system \cite{SantosoPRB2011,GogoiEPL2012}, and its interaction with the underlying substrate can be very interesting to study \cite{GogoiEPL2012}.

In order to study the nature and mechanisms behind these various interface phenomena, it is very crucial to have a thorough understanding of the electronic band structure at the interface and how it differs from those of the parent materials. One way to directly probe this is by measuring the complex dielectric response of the material in a broad energy range \cite{SantosoPRB2011,RusydiPRB2008,MajidiPRB2011,AsmaraJAP}. For example, a combination of spectroscopic ellipsometry (0.5 - 5.6 eV) and ultraviolet - vacuum ultraviolet (UV-VUV) reflectivity (3.7 - 35 eV) can be used to obtain the reflectivity of the material in the broad range of 0.5 - 35 eV \cite{SantosoPRB2011,RusydiPRB2008,MajidiPRB2011,AsmaraJAP}. The broad photon energy range is crucial in order to yield a stabilized Kramers-Kronig analysis of the reflectivity data (explained in details later), so that the correct complex dielectric function $\varepsilon(\omega)=\varepsilon_{1}(\omega)+i\varepsilon_{2}(\omega)$ of the material can be extracted reliably from the reflectivity. This technique has been proven to be important in the study of a wide variety of materials, ranging from manganites, graphene, and oxides such as SrTiO$_{3}$ \cite{SantosoPRB2011,RusydiPRB2008,MajidiPRB2011,AsmaraJAP,VanBenthemJAP2001}.

One important difference between the study of interfaces with that of bulk materials is the fact that the interface is buried under one or more layers of parent materials. Thus, any technique intended to be used in the study of interfaces has to be able to probe the buried interface without disturbing the parent materials surrounding it. For example, in LaAlO$_{3}$/SrTiO$_{3}$ (LAO/STO) heterostructure, the LaAlO$_{3}$ film thickness is typically in the order of 1 - 4 nm, which means that the interface is buried also at 1 - 4 nm below the surface. In the optical reflectivity setup described above, the photon penetration depth is found to be in the order of 10 - 40 nm, which is more than sufficient to probe the buried interface of LaAlO$_{3}$/SrTiO$_{3}$. In general, this is also applicable to other multilayered systems as long as the depth at which the interface is buried does not exceed the penetration depth of the photon.

In this paper, we discuss various aspects of reflectivity and dielectric function analysis of both bulk and multilayered materials. Especially for the multilayer analysis, we present a self-consistent iteration procedure, a versatile method to extract and separate the complex dielectric function of each individual layer of a multilayered system. As the case study, we apply this method to LaAlO$_{3}$/SrTiO$_{3}$ heterostructure \cite{OhtomoNature2004}, in which we are able to separate the effects of the interface from the LaAlO$_{3}$ film and the SrTiO$_{3}$ substrate. Our method can be applied to other multilayered systems with various numbers of layers.

\section{General analysis of optics data of bulk materials}

We first discuss the spectroscopic ellipsometry (SE), which can cover a photon energy range of 0.5 - 5.6 eV. SE is a self-normalizing technique to determine the complex element of dielectric tensor from a single measurement without the need to perform a Kramers-Kronig transformation, making it free from any ambiguities that are related to the normalization of conventional reflectivity results \cite{RauerRSI2005}. The raw data measured SE is expressed in terms of $\Psi$ and $\Delta$, which are defined as \cite{Fujiwara2007}
\begin{equation}\label{eq:eq1}
    \rho \equiv \tan \Psi \exp(i\Delta) \equiv \frac{r_{p}}{r_{s}},
\end{equation}
where $\rho$ is the ratio between $r_{p}$ and $r_{s}$, the reflection coefficients of p- (parallel to the plane of incident) and s- (perpendicular to the plane of incident) polarized light, respectively. From the Fresnel equations, $r_{p}$ and $r_{s}$ are defined as
\begin{equation}\label{eq:eq2}
    r_{p}^{ij} = \frac{n_{j} \cos \theta_{i} - n_{i} \cos \theta_{j}}{n_{j} \cos \theta_{i} + n_{i} \cos \theta_{j}}
\end{equation}
and
\begin{equation}\label{eq:eq3}
    r_{s}^{ij} = \frac{n_{i} \cos \theta_{i} - n_{j} \cos \theta_{j}}{n_{i} \cos \theta_{i} + n_{j} \cos \theta_{j}}.
\end{equation}
Here, $n$ and $\theta$ represent the refraction index and angle of incident from the surface normal, respectively. The $i$ and $j$ indices represent the two materials involved in the photon propagation.

From $n$, the complex dielectric function $\varepsilon(\omega)$ can be obtained using
\begin{equation}\label{eq:eq4}
    \sqrt{\varepsilon(\omega)}=n(\omega),
\end{equation}
where $\omega$ is the photon frequency. The $\varepsilon(\omega)$ obtained using Eq.~\ref{eq:eq4} can then be converted back to reflection coefficients using Eqs.~\ref{eq:eq2} and \ref{eq:eq3}. The $\tan \Psi$ and $\Delta$ are essentially \emph{ratios} of the intensities (for $\tan \Psi$) and phases (for $\Delta$) of the reflection coefficient of the p- and s-polarized lights (Eq.~\ref{eq:eq1}), which makes them (and any quantities derived from them, including reflectivity) \emph{self-normalized}. This is an important attribute of SE, since the converted self-normalized reflectivity can be used to normalize the reflectivity data obtained using other methods, such as the UV-VUV reflectivity technique.

\begin{figure}
\includegraphics[width=3.4in]{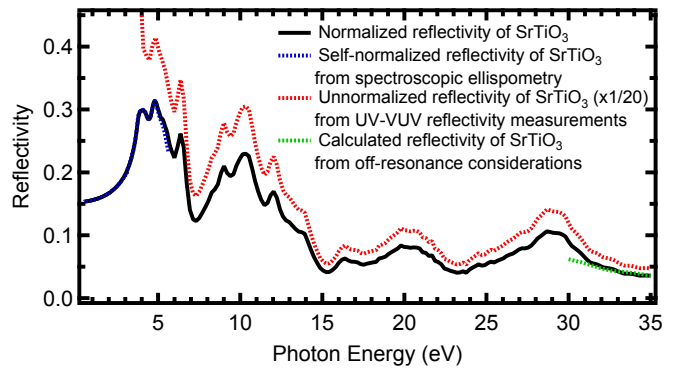}
\caption{The normalized high-energy reflectivity (0.5 - 35 eV) of SrTiO$_{3}$ is compared to the self-normalized reflectivity obtained from spectroscopic ellipsometry (0.5 - 5.6 eV), the unnormalized UV-VUV reflectivity (3.7 - 35 eV) from the UV-VUV reflectivity measurements (scaled down by 20$\times$ to fit the graph), and the calculated reflectivity from off-resonance considerations ($>$ 30 eV). To obtain the normalized reflectivity in the full range of 0.5 - 35 eV, the unnormalized UV-VUV reflectivity is further scaled down to match the spectroscopic ellipsometry and the off-resonance data, and then the three data are appended together. The raw data are reproduced with permission from T. C. Asmara {\it et al.}, Nat. Commun. {\bf 5}, 3663 (2014) \cite{AsmaraNatComm}. Copyright 2014 by Nature Publishing Groups (NPG).}
\label{fig:fig1}
\end{figure}

Next, we discuss the analysis of the UV-VUV reflectivity data (3.7 - 35 eV) of bulk materials. In optics, a material can be considered as a bulk material when its thickness, $d$, is more than five times the photon penetration depth, $D$, (see Section IX for detailed discussion of $D$), $i.e.$ $d>5D$ \cite{Fujiwara2007}. Due to the self-normalized nature of SE, the SE-derived reflectivity can be used to normalize the UV-VUV reflectivity at the low energy side within the range of 3.7 - 5.6 eV. Furthermore, the high energy part ($>$ 30 eV) is normalized using calculations based on off-resonance scattering considerations according to \cite{HenkeADNDT1993}
\begin{equation}\label{eq:eq5}
    r = i \frac{r_{0} \lambda}{\sin \theta} F(\theta)P_{f}(2 \theta),
\end{equation}
where $r$ is the reflection coefficient, $r_{0}$ is the classical electron radius ($e^{2}/mc^{2}$), $\lambda$ is the photon wavelength, $P_{f}(\theta)$ is the polarization factor (equal to unity for s-polarized light and equal to $\cos \theta$ for p-polarized light), and $F(\theta)$ is the structure factor per unit area given by
\begin{equation}\label{eq:eq6}
    F(\theta) = \sum_{q} n_{q} f_{q} \exp (i\frac{4 \pi z_{q}}{\lambda} \sin \theta).
\end{equation}
The summation is performed over the different types of atoms on a particular atomic plane on which the light is incident, with $n_{q}$ denotes the number of atoms of type $q$ in that particular plane, $f_{q}$ denotes the tabulated atomic form factor corresponding to that atom $q$, and $z_{q}$ denotes the direction vector normal to the plane in question. From the above step, normalized reflectivity in the range of 0.5 - 35 eV can be obtained. As an illustration for the normalization procedure, Figure~\ref{fig:fig1} shows the normalized high-energy reflectivity (0.5 - 35 eV) of SrTiO$_{3}$ as compared to the self-normalized reflectivity obtained from SE (0.5 - 5.6 eV), the unnormalized UV-VUV reflectivity (3.7 - 35 eV), and the off-resonance considerations ($>$ 30 eV). (The details of the experimental procedures used to obtain the data are discussed in Section IV.)

From the above step, normalized reflectivity in the range of 0.5 - 35 eV can be obtained. For isotropic bulk materials, the $\varepsilon(\omega)$ can be extracted from the normalized UV-VUV reflectivity using Kramers-Kronig analysis \cite{KramersNature1926,KronigJOSA1926,TollPR1956,JahodaPR1957,NilssonPkM1969,Landau1975,Dressel2002,KuzmenkoRSI2005}. The procedure is as following. The reflection coefficient and phase difference between the reflected and incident light, $\varphi$, are related through the Kramers-Kronig transformation according to
\begin{equation}\label{eq:eq7}
    r(\omega)=\sqrt{R(\omega)}\exp(i\varphi(\omega))
\end{equation}
and
\begin{equation}\label{eq:eq8}
    \varphi(\omega)=-\frac{\omega}{\pi}P\int_{0}^{\infty}\frac{\ln R(x)}{x^{2}-\omega^{2}}dx+\varphi(0),
\end{equation}
where $R=|r|^{2}$ is the reflectivity and $P$ is the Cauchy principal value. From here, the refractive index $n$ and extinction coefficient $k$ can be extracted from reflectivity using
\begin{equation}\label{eq:eq9}
    n=\frac{1-R}{1+R-2\sqrt{R\cos \varphi}}
\end{equation}
and
\begin{equation}\label{eq:eq10}
    k=\frac{2R\sin \varphi}{1+R-2\sqrt{R\cos \varphi}}.
\end{equation}
Finally, the real ($\varepsilon_{1}$) and imaginary ($\varepsilon_{2}$) parts of complex dielectric function can be obtained from $n$ and $k$ via
\begin{equation}\label{eq:eq11}
    \varepsilon_{1}=n^{2}-k^{2}
\end{equation}
and
\begin{equation}\label{eq:eq12}
    \varepsilon_{2}=2nk.
\end{equation}
Like $R$ and $\varphi$, $\varepsilon_{1}$ and $\varepsilon_{2}$ are also related through the Kramers-Kronig relationship according to
\begin{equation}\label{eq:eq13}
    \varepsilon_{1}(\omega)-1=\frac{2}{\pi}P\int_{0}^{\infty}\frac{x\varepsilon_{2}(x)}{x^{2}-\omega^{2}}dx
\end{equation}
and
\begin{equation}\label{eq:eq14}
    \varepsilon_{2}(\omega)=-\frac{2\omega}{\pi}P\int_{0}^{\infty}\frac{\varepsilon_{1}(x)}{x^{2}-\omega^{2}}dx+\frac{4\pi\sigma_{DC}}{\omega},
\end{equation}
where $\sigma_{DC}$ is the DC conductivity.

The Kramers-Kronig analysis can be either done directly through function inversion ($i.e.$ by directly using Eqs.~\ref{eq:eq7}-\ref{eq:eq12}), numerical approximation, or through fitting. In this paper, the analysis is done by fitting \cite{KuzmenkoRSI2005} using the Kramers-Kronig-transformable Drude-Lorentz oscillators according to
\begin{equation}\label{eq:eq15}
    \varepsilon(\omega)=\varepsilon_{\infty}+\sum_{k}\frac{\omega_{p,k}^{2}}{\omega_{0,k}^{2}-\omega^{2}-i\Gamma_{k}\omega}.
\end{equation}
The high-frequency dielectric constant is denoted by $\varepsilon_{\infty}$; $\omega_{p,k}$, $\omega_{0,k}$, and $\Gamma_{k}$ are the plasma frequency, the transverse frequency (eigenfrequency), and the line width (scattering rate) of the $k$-th oscillator, respectively. Since the energy range involved is very broad (covering 0.5 - 35 eV), the analysis yields a stabilized Kramers-Kronig transformation.

\section{General overview of analysis of multilayered materials}

If the material is in isotropic bulk form, the Kramers-Kronig analysis is straightforward, since the obtained reflectivity only depends on $\varepsilon(\omega)$ of one material. However if the material is composed of several layers ($i.e.$ a multilayer), the analysis becomes more complex due to the interference between the light reflected from the surface and those reflected from the interface(s).  For example, according to analysis of wave propagation in a stratified medium, the reflection coefficient of a thin film on a substrate ($i.e.$ a two-layered material) has the following form \cite{Born2003},
\begin{equation}\label{eq:eq16}
    r_{amb,multi}=\frac{r_{amb,film} + r_{film,subs} e^{i2 \delta_{film}}}{1+ r_{amb,film} r_{film,subs} e^{i2 \delta_{film}}},
\end{equation}
where
\begin{equation}\label{eq:eq17}
    \delta_{i}=\frac{2\pi d_{i}}{\lambda}n_{i}\cos \theta_{i}.
\end{equation}
Here, the subscripts $amb$, $multi$, $film$, and $subs$ represent the ambient, the multilayer, the thin film, and the substrate, respectively, which are the various materials involved in the propagation of the photon.

In other words, for a multilayered system the obtained reflection coefficient, reflectivity ($R_{amb,multi}=|r_{amb,multi}|^{2}$), and, via Eq.~\ref{eq:eq1}, $\rho$, $\Psi$, and $\Delta$ depend on the $\varepsilon(\omega)$ of both films and substrate, along with the thickness of the film and the angle of incidence \cite{ArwinTSF1984,ArwinTSF1986},
\begin{equation}\label{eq:eq18}
\begin{split}
    R_{amb,multi}=R(\varepsilon_{film},\varepsilon_{subs},d_{film},\theta),\\
    \rho_{amb,multi}=\rho(\varepsilon_{film},\varepsilon_{subs},d_{film},\theta),
\end{split}
\end{equation}
and a detailed analysis is required to separate them. In this paper, we discuss various aspects of such analysis, and present the self-consistent iteration procedure, a versatile method to extract and separate the $\varepsilon(\omega)$ of each individual layer of a multilayered system so that they can be further analyzed separately.

If the properties of the substrate are not expected to be significantly affected by the presence of the films, the reflectivity of the bare substrate can be separately measured, and from that its $\varepsilon(\omega)$ can be separately obtained using the general procedure described in Section II. Then, if there is only one film layer, the $\varepsilon(\omega)$ of the film can be straightforwardly obtained from the total reflectivity using Eqs.~\ref{eq:eq2},~\ref{eq:eq3},~\ref{eq:eq16}, and~\ref{eq:eq17}. However, if there are multiple layers of films composed of different materials, and the $\varepsilon(\omega)$ of each material is unknown (or different from their bulk forms), then the analysis becomes more complicated. This is because there are several unknowns but only one equation (Eq.~\ref{eq:eq16}) \cite{ArwinTSF1986}, which prevents a straightforward mathematical solution. The same problem also occurs if the properties of the substrate are affected by the presence of the films. For example, if parts of the substrate near the interface become modified due to the presence of the films, then in the analysis the interface needs to be treated as an effective additional layer. So, even if the system is composed of only one layer of thin film on top of a substrate (an apparent two-layered system), it needs to be treated as if it was a three-layered one due to the presence of the interface layer.

\section{Case study: LAO/STO heterostructure}

\begin{figure}
\includegraphics[width=3.4in]{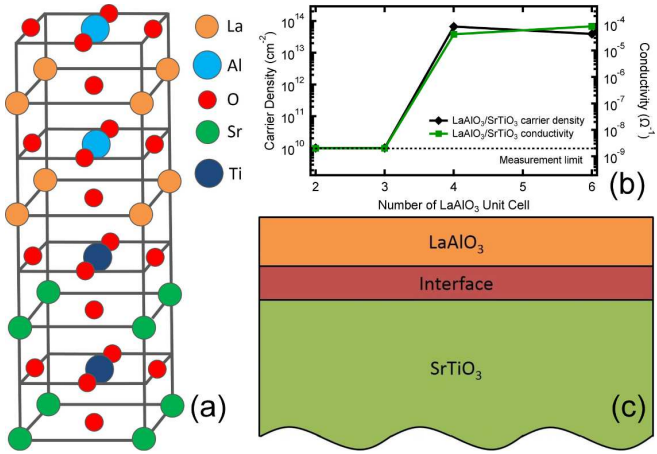}
\caption{(a) Crystal structure of LaAlO$_{3}$/SrTiO$_{3}$. (b) Transport measurement of LaAlO$_{3}$/SrTiO$_{3}$, showing the LaAlO$_{3}$-thickness-dependent metal-insulator transition. The raw data are reproduced with permission from T. C. Asmara {\it et al.}, Nat. Commun. {\bf 5}, 3663 (2014) \cite{AsmaraNatComm}. Copyright 2014 by Nature Publishing Groups (NPG). (c) Multilayer consideration of LaAlO$_{3}$/SrTiO$_{3}$.}
\label{fig:fig2}
\end{figure}

One example is the LaAlO$_{3}$/SrTiO$_{3}$ heterostructure (Figure~\ref{fig:fig2} (a)), which is also the system that will be used as the case study throughout this paper. When thin film of LaAlO$_{3}$ is deposited on SrTiO$_{3}$, a conducting quasi two-dimensional electron gas (2DEG) with high mobility and carrier density emerges at the interface \cite{OhtomoNature2004}. Interestingly, this quasi-2DEG only emerges when the LaAlO$_{3}$ film reaches a certain critical thickness, usually 4 unit cells (uc) or more \cite{ThielScience2006}. Below the critical thickness, the system remains insulating. Furthermore, the interface is also found to exhibit magnetism \cite{BrinkmanNatMater2007,SeriPRB2009,AriandoNatComm2011,KaliskyNatComm2012,LeeNatMat2013} and superconductivity \cite{ReyrenScience2007,CavigliaNature2008,BenShalomPRL2010}, and two-dimensional coexistence of both has even been observed \cite{DikinPRL2011,LiNatPhys2011,BertNatPhys2011}. These observations are very remarkable considering LaAlO$_{3}$ and SrTiO$_{3}$ are non-magnetic insulators in their bulk state \cite{AsmaraNatComm}. Thus, the understanding of electronic band structure at the LaAlO$_{3}$/SrTiO$_{3}$ interface is very crucial to reveal the nature and mechanisms of these interesting phenomena.

For the case study, four samples of LaAlO$_{3}$/SrTiO$_{3}$ with varying thicknesses of LaAlO$_{3}$ (2, 3, 4, and 6 uc, respectively) are prepared using techniques described elsewhere \cite{AsmaraNatComm,AriandoNatComm2011}. From transport measurements, it is known that the 2 and 3 uc samples are insulating with carrier density and conductivity below the measurement limit, while the 4 and 6 uc ones are conducting with carrier density of $4 - 6 \times 10^{13}$ cm$^{-2}$ and conductivity of $4 - 8 \times 10^{-5}$ $\Omega^{-1}$, consistent with previous results \cite{ThielScience2006,BrinkmanNatMater2007,AriandoNatComm2011,ReyrenScience2007}(Figure~\ref{fig:fig2} (b)). Bulk LaAlO$_{3}$ and bulk SrTiO$_{3}$ samples are also prepared for comparison.

The reflectivity is obtained using a combination of SE (0.5 - 5.6 eV), and UV-VUV reflectivity (3.7 - 35 eV) \cite{SantosoPRB2011,RusydiPRB2008,MajidiPRB2011,AsmaraJAP}. The details of the optical measurements are as follow. The SE measurements are performed in the spectral range between 0.5 and 5.6 eV by using an SE 850 ellipsometer at room temperature \cite{RauerRSI2005}. Three different incident angles of 60$^{\circ}$, 70$^{\circ}$, and 80$^{\circ}$ from the sample normal are needed, and the incident light is 45$^{\circ}$ linearly polarized from the plane of incident. For reflectivity measurements in the high-energy range between 3.7 and 35 eV we use the superlumi beamline at the DORIS storage ring of HASYLAB (DESY) \cite{ZimmererNIMPRA1991}. The incoming photon is incident at the angle of 17.5$^{\circ}$ from the sample normal with linear polarization parallel to the sample surface. The sample chamber is outfitted with a gold mesh to measure the incident photon flux after the slit of the monochromator. The measurements are performed in ultrahigh vacuum environment (chamber pressure of $10^{-9}$ mbar) at room temperature. The obtained UV-VUV reflectivity data is calibrated by comparing it with the luminescence yield of sodium salicylate (NaC$_{7}$H$_{5}$O$_{3}$) and the gold mesh.

For the nearly-isotropic bulk LaAlO$_{3}$ and bulk SrTiO$_{3}$, the reflectivity normalization along with the extraction of $\varepsilon(\omega)$ from the normalized reflectivity can be performed using the general procedure described in Section II. (The normalized reflectivity and extracted $\varepsilon(\omega)$ of both are shown later in Figure~\ref{fig:fig6}.) On the other hand, the analysis of LaAlO$_{3}$/SrTiO$_{3}$ is not as straightforward, due to its heterostructure nature as well as the presence of the conducting layer at its interface. For this reason, a multilayer consideration based on a boundary analysis of Fresnel equation needs to be taken into account in analyzing the SE data and the UV-VUV reflectivity of LaAlO$_{3}$/SrTiO$_{3}$. In this multilayer analysis, LaAlO$_{3}$/SrTiO$_{3}$ (particularly the conducting samples) consists of three layers: the LaAlO$_{3}$ film layer at the top, the bulk SrTiO$_{3}$ layer at the bottom and an interface layer sandwiched between the two at the middle, representing the quasi-2DEG at the interface (Figure~\ref{fig:fig2} (c)), consistent with previous observation using cross-sectional conducting tip atomic force microscopy \cite{BasleticNatMat2008}.

According to analysis of wave propagation in a stratified medium, the reflection coefficient of a three-layer system like LaAlO$_{3}$/SrTiO$_{3}$ can be expressed through Fresnel equations as \cite{Born2003}
\begin{widetext}
\begin{equation}\label{eq:eq19}
    r_{amb,multi}=\frac { r_{amb,fLAO} + r_{fLAO,int} e^{i2 \delta_{fLAO}} + r_{amb,fLAO} r_{fLAO,int} r_{int,STO} e^{i2 \delta_{int}} + r_{int,STO} e^{i2 (\delta_{fLAO} + \delta_{int})}} {1 + r_{amb,fLAO} r_{fLAO,int} e^{i2 \delta_{fLAO}} + r_{fLAO,int} r_{int,STO} e^{i2 \delta_{int}} + r_{amb,fLAO} r_{int,STO} e^{i2 (\delta_{fLAO} + \delta_{int})}},
\end{equation}
\end{widetext}
where the subscripts $fLAO$, $int$, and STO represent the LaAlO$_{3}$ film, the interface layer, and the SrTiO$_{3}$ substrate, respectively. Thus, the reflectivity (and, by extension via Eq.~\ref{eq:eq1}, $\rho$, $\Psi$, and $\Delta$) of LaAlO$_{3}$/SrTiO$_{3}$ contains mixed information from all three constituent layers \cite{ArwinTSF1984,ArwinTSF1986},
\begin{equation}\label{eq:eq20}
\begin{split}
    R_{amb,multi}=R(\varepsilon_{fLAO},\varepsilon_{int},\varepsilon_{STO},d_{fLAO},d_{int},\theta),\\
    \rho_{amb,multi}=\rho(\varepsilon_{fLAO},\varepsilon_{int},\varepsilon_{STO},d_{fLAO},d_{int},\theta).
\end{split}
\end{equation}

This makes the extraction of the $\varepsilon(\omega)$ of individual layer non-trivial, because there are too many unknown factors involved. Since $\varepsilon(\omega)$ of the bulk SrTiO$_{3}$ substrate, $\varepsilon_{STO}$, can be measured independently and LaAlO$_{3}$ film thickness, $d_{fLAO}$, of each sample are known, Eqs.~\ref{eq:eq1}-\ref{eq:eq4},~\ref{eq:eq17}, and~\ref{eq:eq19} left us with 3 unknown variables: $\varepsilon(\omega)$ of LaAlO$_{3}$ film, $\varepsilon_{fLAO}$, (which might be different from that of bulk LaAlO$_{3}$), $\varepsilon(\omega)$ of interface layer, $\varepsilon_{int}$, and the thickness of the interface layer, $d_{int}$,
\begin{equation}\label{eq:eq21}
\begin{split}
    R_{amb,multi}=R(\varepsilon_{fLAO},\varepsilon_{int},d_{int},\theta),\\
    \rho_{amb,multi}=\rho(\varepsilon_{fLAO},\varepsilon_{int},d_{int},\theta).
\end{split}
\end{equation}
(Note that even though $\varepsilon(\omega)$ has real and imaginary parts, they are connected through the Kramer-Kronig relationship, see Eqs.~\ref{eq:eq13} and~\ref{eq:eq14}.) This poses a challenge, because (assuming there is no change in $\varepsilon_{STO}$ across the samples) there are 3 unknowns but only 1 equation (Eq.~\ref{eq:eq19}), which prevents a straightforward mathematical solution \cite{ArwinTSF1986}. To overcome this problem, it can be noted that the light phase, $\delta$, in Eq.~\ref{eq:eq17} depends mainly on two parameters: the incident angle, $\theta$, (angle-dependent) \cite{IbrahimJOSA1971,SoJOSA1972,HunderiSS1976} and the layer thickness, $d$, (thickness-dependent) \cite{SoJOSA1972}. This means Eq.~\ref{eq:eq17} can be used to diversify Eq.~\ref{eq:eq19} by varying either of these two parameters, so that the number of equations can match the number of unknown variables (in this case, three). (Note that Eq.~\ref{eq:eq19} can also be diversified by varying the ambient within which the measurement is performed \cite{SoJOSA1972,LukesSS1969}, $e.g.$ by purging the measurement chamber with different ambient gas or immersing the setup inside different liquids, however concerns about surface contamination on surface-sensitive samples may make this method less versatile.) This enables us to perform a self-consistent iteration procedure on the reflectivity (and thus also $\rho$, $\Psi$, and $\Delta$ via Eq.~\ref{eq:eq1}) data, so that each unknown variable can be extracted separately.

\section{Angle-dependent iteration procedure}

In SE (0.5 - 5.6 eV), the $\Psi$ and $\Delta$ measurements are done at three different incident angles: 60$^{\circ}$, 70$^{\circ}$, and 80$^{\circ}$ from the sample normal, which results in three sets of $\Psi$ and $\Delta$ data. Since Eq.~\ref{eq:eq1} can be diversified via Eqs.~\ref{eq:eq17} and~\ref{eq:eq19} by varying $\theta$ \cite{IbrahimJOSA1971,SoJOSA1972,HunderiSS1976}, this gives us the three equations necessary to perform an angle-dependent iteration procedure to extract the three unknown variables: $\varepsilon_{fLAO}$, $\varepsilon_{int}$, and $d_{int}$.

\begin{figure}
\includegraphics[width=3.4in]{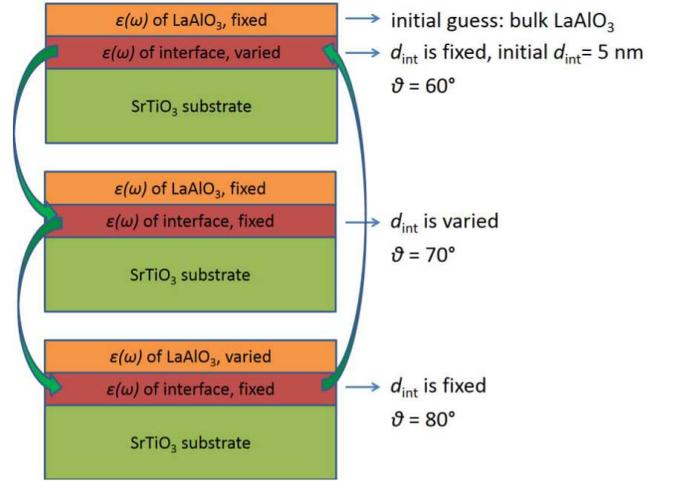}
\caption{Diagram depicting the angle-dependent iteration process.}
\label{fig:fig3}
\end{figure}

As a representative, the iteration for the SE data of the 4 uc LaAlO$_{3}$/SrTiO$_{3}$ sample can be performed as following (Figure~\ref{fig:fig3}). According to previous studies, the thickness of the conducting interface might be around 2 - 10 nm \cite{ReyrenScience2007,BasleticNatMat2008,SiemonsPRL2007,JanickaPRL2009,SingPRL2009,SonPRB2009,DubrokaPRL2010}, so the initial guess for $d_{int}$, $d_{int[0]}$, can be reasonably set as 5 nm. In general, there are two boundary conditions that can be applied when setting the initial guess, and also for confirming the physical validity of the converged value after iteration, of the thickness of films and interfaces:\\
1. The thickness should not be lower than the thickness of 1 uc of the materials. For films, this is the thinnest physical limit for the layer-by-layer deposition, while for interfaces this is to take into account any interface roughness effects.\\
2. The thickness should not be higher than five times the photon penetration depth $D$ (see Section IX for detailed discussion of $D$), since beyond this limit the material is optically considered to be bulk  \cite{Fujiwara2007}.

Meanwhile, the initial guess for $\varepsilon_{fLAO}$ can be set as the same as $\varepsilon(\omega)$ of bulk LaAlO$_{3}$, $\varepsilon_{bLAO}$, which can be obtained independently by measuring it separately. With these two variables fixed, Eq.~\ref{eq:eq1} is fitted into experimental value of $\Psi$ and $\Delta$ measured at $\theta$ = 60$^{\circ}$ using Eq.~\ref{eq:eq15} by appropriately adjusting the Drude-Lorentz oscillators that make up the $\varepsilon(\omega)$ of interface layer \cite{KuzmenkoRSI2005}. At this first step, by expanding Eq.~\ref{eq:eq21} to the first order, the tentative $\varepsilon(\omega)$ of interface after fitting, $\varepsilon_{int[1]}$, can be expressed as \cite{ArwinTSF1984,ArwinTSF1986},
\begin{equation}\label{eq:eq22}
\begin{split}
    \varepsilon_{int[1]}=&\varepsilon_{int} + (d_{int}-d_{int[0]}) \frac {\partial \rho_{amb,multi} / \partial d_{int}} {\partial \rho_{amb,multi} / \partial \varepsilon_{int}}\\
                         &+ (\varepsilon_{fLAO}-\varepsilon_{bLAO}) \frac{\partial\rho_{amb,multi}/\partial\varepsilon_{fLAO}}{\partial\rho_{amb,multi}/\partial\varepsilon_{int}}.
\end{split}
\end{equation}

It can be seen that at this step, $\varepsilon_{int[1]}$ deviates from the actual value of $\varepsilon_{int}$ due to the still-improper values of $d_{int}$ and $\varepsilon_{fLAO}$. To simplify the notation, generalized addition and subtraction operators, $\oplus$ and $\ominus$, respectively, can be introduced to represent the correlation effects of $d_{int}$ and $\varepsilon_{fLAO}$ on $\varepsilon_{int[1]}$ such that,
\begin{equation}\label{eq:eq23}
    \varepsilon_{int[1]}=\varepsilon_{int} \oplus (\varepsilon^{d}_{int}\ominus\varepsilon^{d}_{int[0]}) \oplus (\varepsilon_{fLAO}\ominus\varepsilon_{bLAO}).
\end{equation}
Here, $\varepsilon^{d}_{int}$ and $\varepsilon^{d}_{int[0]}$ are introduced to represent the correlation effects that $d_{int}$ and $d_{int[0]}$ have on $\varepsilon_{int[1]}$, respectively. For example, to the first order, $\varepsilon^{d}_{int}$ can be expressed according to Eq.~\ref{eq:eq22} as,
\begin{equation}\label{eq:eq24}
   \varepsilon^{d}_{int}=\frac {\partial \rho_{amb,multi} / \partial d_{int}} {\partial \rho_{amb,multi} / \partial \varepsilon_{int}} d_{int}.
\end{equation}
Another advantage of these generalized operators notations is that they also allows the higher orders of Eq.~\ref{eq:eq21} expansion to be implicitly included in Eq.~\ref{eq:eq23}, meaning that Eq.~\ref{eq:eq23} can be taken as the generalized form of Eq.~\ref{eq:eq22}. Thus, due to their convenience, these operators shall be used throughout this paper.

After the first step described above, the newly-fitted $\varepsilon_{int[1]}$ is in turn fixed, and $d_{int}$ is appropriately adjusted so that Eq.~\ref{eq:eq1} can now be fitted into experimental value of $\Psi$ and $\Delta$ measured at $\theta$ = 70$^{\circ}$. Here, the tentative value of $d_{int}$, $d_{int[1]}$, can be expressed as,
\begin{equation}\label{eq:eq25}
    d_{int[1]}=d_{int} \oplus (d^{\varepsilon}_{fLAO} \ominus d^{\varepsilon}_{bLAO}) \oplus (d^{\varepsilon}_{int}\ominus d^{\varepsilon}_{int[1]}).
\end{equation}
Again, $d^{\varepsilon}_{fLAO}$, $d^{\varepsilon}_{bLAO}$, $d^{\varepsilon}_{int}$, and $d^{\varepsilon}_{int[1]}$ are introduced to represent the correlation effects that $\varepsilon_{fLAO}$, $\varepsilon_{bLAO}$, $\varepsilon_{int}$, and $\varepsilon_{int[1]}$ have on $d_{int[1]}$, respectively. As a representative, $d^{\varepsilon}_{fLAO}$ can be expressed to the first order as,
\begin{equation}\label{eq:eq26}
   d^{\varepsilon}_{fLAO}=\frac {\partial \rho_{amb,multi} / \partial \varepsilon_{fLAO}} {\partial \rho_{amb,multi} / \partial d_{int}} \varepsilon_{fLAO}.
\end{equation}

\begin{figure}
\includegraphics[width=3.4in]{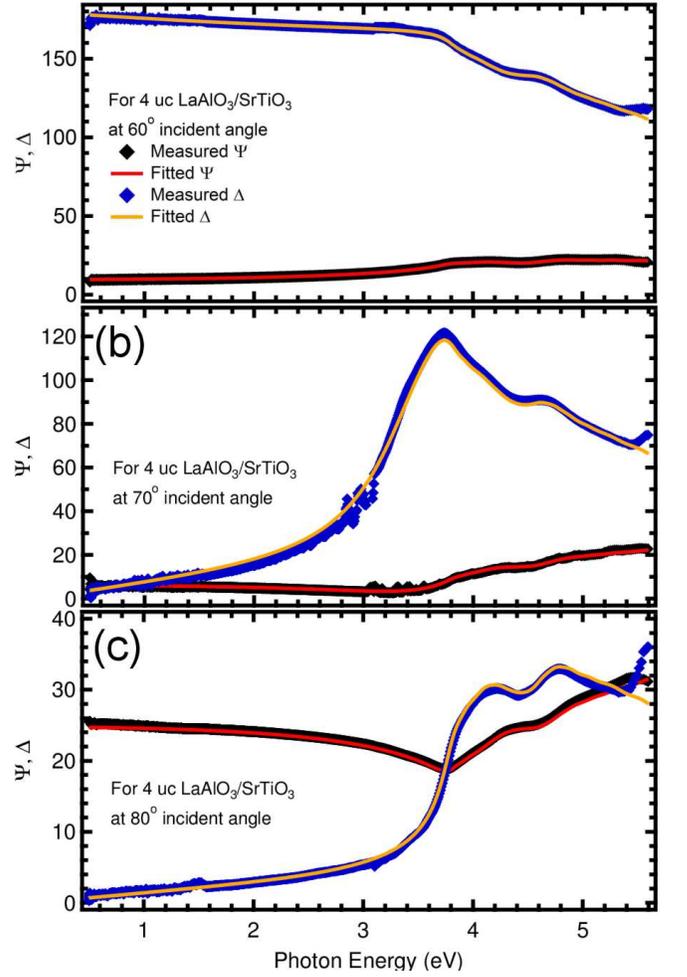}
\caption{Comparison between the experimentally-measured $\Psi$ and $\Delta$ of 4 uc LaAlO$_{3}$/SrTiO$_{3}$ and their fitted values after iteration for all three incident angles. (a) For 60$^{\circ}$ incident angle. (b) For 70$^{\circ}$ incident angle. (c) For 80$^{\circ}$ incident angle. The fitted values match the measured $\Psi$ and $\Delta$ very well for all three incident angles, confirming the stability of the iteration. The raw data are reproduced with permission from T. C. Asmara {\it et al.}, Nat. Commun. {\bf 5}, 3663 (2014) \cite{AsmaraNatComm}. Copyright 2014 by Nature Publishing Groups (NPG).}
\label{fig:fig4}
\end{figure}

Then, the newly-adjusted $d_{int}[1]$ is also fixed (along with the previously-fitted $\varepsilon_{int[1]}$), and Eq.~\ref{eq:eq1} is fitted into experimental value of $\Psi$ and $\Delta$ measured at $\theta$ = 80$^{\circ}$ using Eq.~\ref{eq:eq15} by appropriately adjusting the Drude-Lorentz oscillators that make up the $\varepsilon(\omega)$ of LaAlO$_{3}$ film layer. Here, the tentative $\varepsilon(\omega)$ of LaAlO$_{3}$ film, $\varepsilon_{fLAO[1]}$, can be expressed as,
\begin{equation}\label{eq:eq27}
    \varepsilon_{fLAO[1]}=\varepsilon_{fLAO} \oplus (\varepsilon_{int}\ominus\varepsilon_{int[1]}) \oplus (\varepsilon^{d}_{int}\ominus\varepsilon^{d}_{int[1]}).
\end{equation}
After that, the process is repeated by going back to $\Psi$ and $\Delta$ values measured at $\theta$ = 60$^{\circ}$ and subsequently cycling through the incident angles, fitting only one variable at each step while keeping the other two fixed.

Convergence is reached when, at a certain step $N$, $\varepsilon_{fLAO[N]}$, $\varepsilon_{int[N]}$, and $d_{int}[N]$ can satisfy Eq.~\ref{eq:eq1} for all three incident angles, as shown in Figure~\ref{fig:fig4}. At this point, $ \varepsilon_{fLAO[N]} \rightarrow \varepsilon_{fLAO}$, $\varepsilon_{int[N]} \rightarrow \varepsilon_{int}$, $d_{int[N]} \rightarrow d_{int}$, and the correlation effects between these three parameters are minimized (see Section VI for a more rigorous treatment of this convergence condition). In other words, the iteration results form a universal fitting that can match the data from all incident angles. The iteration thus results in the converged values of $\varepsilon_{fLAO}$, $\varepsilon_{int}$, and $d_{int}$, as shown in Figure~\ref{fig:fig14} later. Along with the already-known $\varepsilon_{STO}$ and $d_{fLAO}$, these quantities can be converted to reflectivity in the 0.5 - 5.6 eV range using Eqs.~\ref{eq:eq2} - \ref{eq:eq4}, \ref{eq:eq17}, and \ref{eq:eq19}, which then can be used to normalize the UV-VUV reflectivity.

\begin{figure}
\includegraphics[width=3.4in]{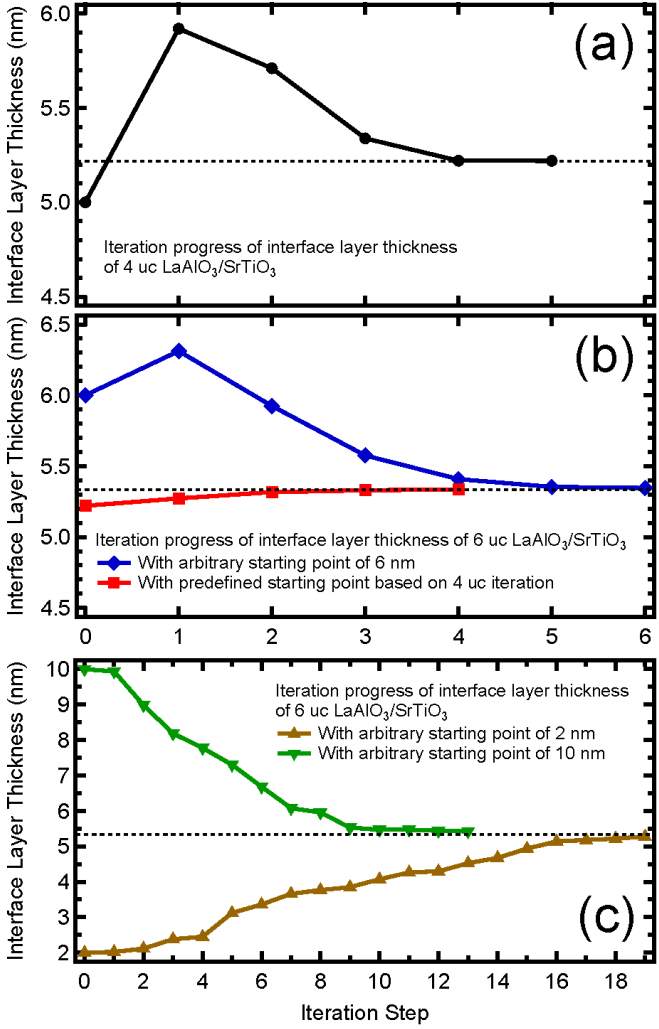}
\caption{(a) Iteration progress of the interface layer thickness for 4 uc LaAlO$_{3}$/SrTiO$_{3}$. (b) Iteration progress of the interface layer thickness for 6 uc LaAlO$_{3}$/SrTiO$_{3}$, showing the comparison between the arbitrary and predefined starting points of 6 nm and 5.2 nm, respectively. (c) Iteration progress of the interface layer thickness for 6 uc LaAlO$_{3}$/SrTiO$_{3}$, showing the comparison between the arbitrary starting points of 2 nm and 10 nm, respectively.}
\label{fig:fig5}
\end{figure}

Figure~\ref{fig:fig5} (a) illustrates the iteration process of the 4 uc LaAlO$_{3}$/SrTiO$_{3}$ by showing the evolution of $d_{int}$ through each iteration step. As the iteration progresses, the value of $d_{int}$ slowly approaches a distinct asymptotic value, and at step 5 it finally converges into $\sim$5.2 nm.

\begin{figure}
\includegraphics[width=3.4in]{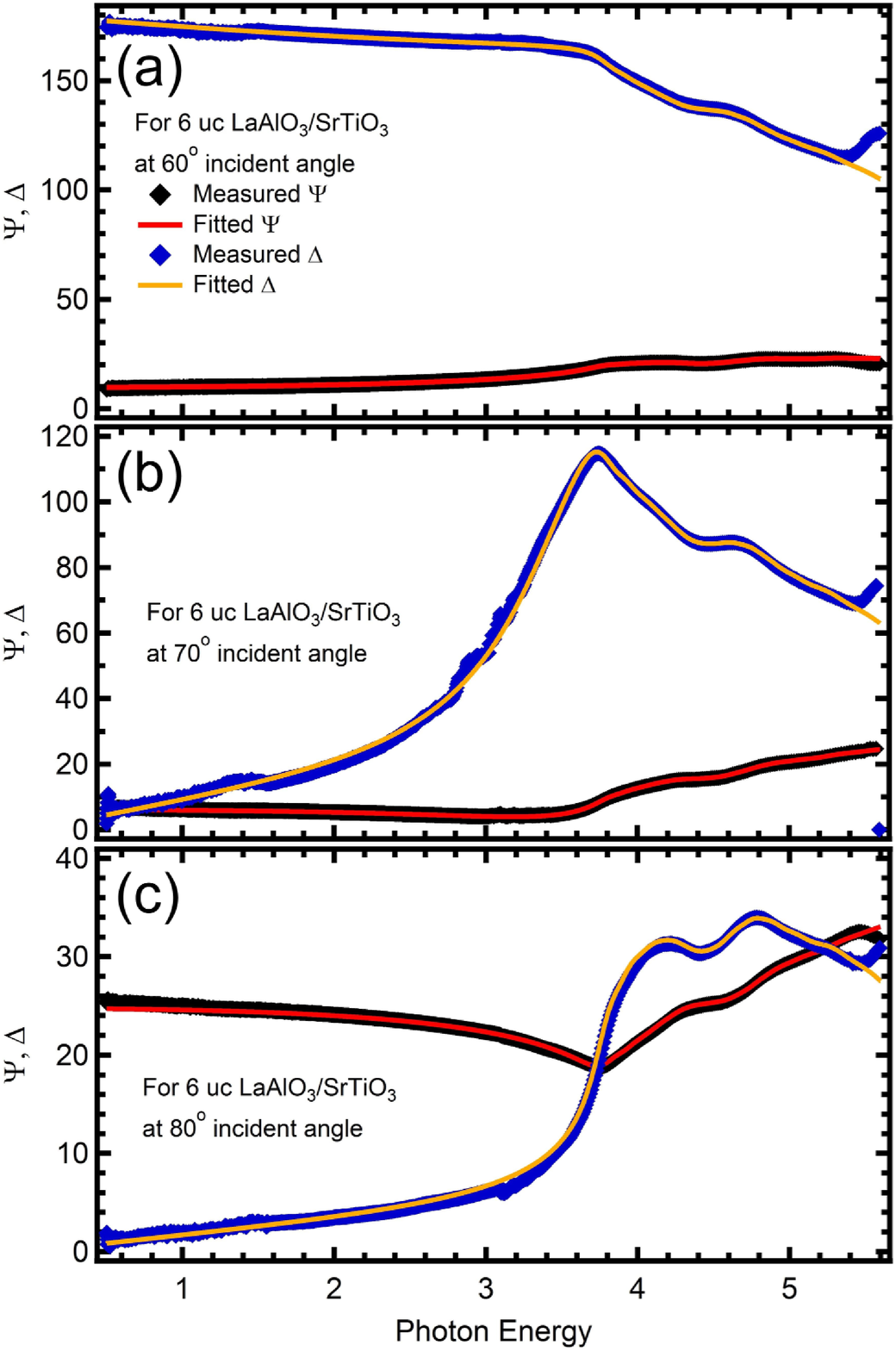}
\caption{Comparison between the experimentally-measured $\Psi$ and $\Delta$ of 6 uc LaAlO$_{3}$/SrTiO$_{3}$ and their fitted values after iteration for all three incident angles. (a) For 60$^{\circ}$ incident angle. (b) For 70$^{\circ}$ incident angle. (c) For 80$^{\circ}$ incident angle. The fitted values match the measured $\Psi$ and $\Delta$ very well for all three incident angles, confirming the stability of the iteration. The raw data are reproduced with permission from T. C. Asmara {\it et al.}, Nat. Commun. {\bf 5}, 3663 (2014) \cite{AsmaraNatComm}. Copyright 2014 by Nature Publishing Groups (NPG).}
\label{fig:fig6}
\end{figure}

For 6 uc LaAlO$_{3}$/SrTiO$_{3}$, the iteration process can be performed similarly, since the only difference is $d_{fLAO}$, which is known and can be appropriately adjusted using Eq.~\ref{eq:eq17}. Figure~\ref{fig:fig6} shows the fitted $\Psi$ and fitted $\Delta$ after convergence that match the measured $\Psi$ and measured $\Delta$ of 6 uc LaAlO$_{3}$/SrTiO$_{3}$. The iteration progress of $d_{int}$ for 6 uc LaAlO$_{3}$/SrTiO$_{3}$ is shown in Figure~\ref{fig:fig5} (b). In this case, the initial guess for $d_{int}$ is set to be 6 nm, and the final converged value is found to be $\sim$5.3 nm, very close to the 4 uc value of $\sim$5.2 nm. This indicates that the properties of 4 and 6 uc LaAlO$_{3}$/SrTiO$_{3}$ are very similar, and any apparent differences in $\Psi$, $\Delta$, and reflectivity values between the two samples are mainly caused by the difference in $d_{fLAO}$. In fact, because of this, since from 4 uc iteration the converged values for $\varepsilon_{fLAO}$, $\varepsilon_{int}$, and $d_{int}$ of 4 uc LaAlO$_{3}$/SrTiO$_{3}$ are already obtained, those values can also be used as the initial guess for the 6 uc LaAlO$_{3}$/SrTiO$_{3}$ iteration. It can be seen from Figure~\ref{fig:fig5} (b) that with those better starting points, the iteration process can be simplified and convergence can be achieved with fewer steps, while still reaching the same converged value of $d_{int}$ $\approx$ 5.3 nm. Further tests of the stability of the iteration process are also done by setting the initial guess for $d_{int}$ to be 2 nm and 10 nm, respectively. From Figure~\ref{fig:fig5} (c), it can be seen that both iterations are indeed able to converge to the same $d_{int}$ value of $\sim$5.3 nm, although they need considerably more steps to converge because the initial guesses deviate a lot more from the converged value. These results confirm the self-consistency of the iteration process, showing that even if it starts with different initial guesses, the iteration does eventually converge into the same final results.

\begin{figure}
\includegraphics[width=3.4in]{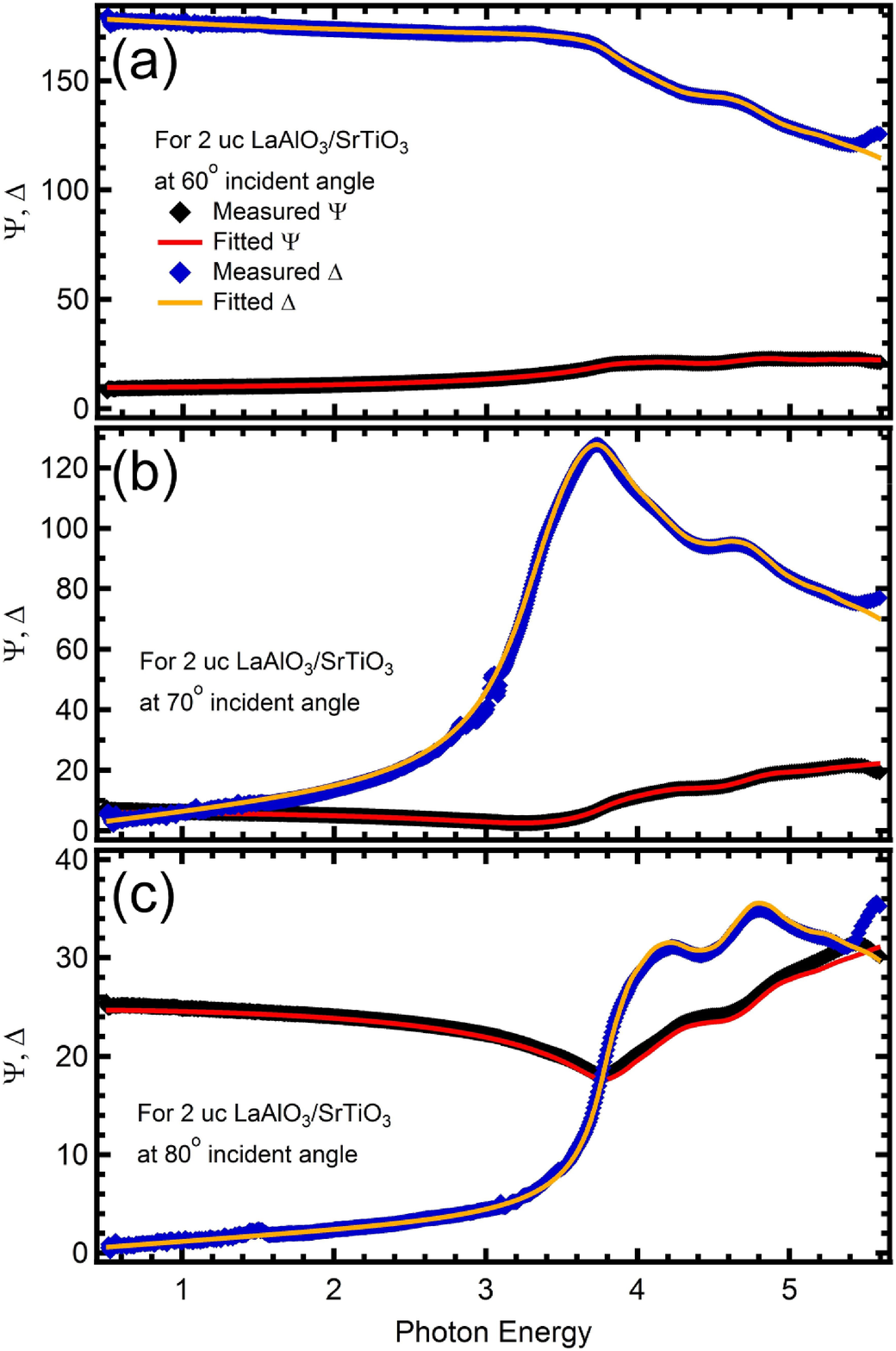}
\caption{Comparison between the experimentally-measured $\Psi$ and $\Delta$ of 2 uc LaAlO$_{3}$/SrTiO$_{3}$ and their fitted values after iteration for all three incident angles. (a) For 60$^{\circ}$ incident angle. (b) For 70$^{\circ}$ incident angle. (c) For 80$^{\circ}$ incident angle. The fitted values match the measured $\Psi$ and $\Delta$ very well for all three incident angles, confirming the stability of the iteration. The raw data are reproduced with permission from T. C. Asmara {\it et al.}, Nat. Commun. {\bf 5}, 3663 (2014) \cite{AsmaraNatComm}. Copyright 2014 by Nature Publishing Groups (NPG).}
\label{fig:fig7}
\end{figure}

\begin{figure}
\includegraphics[width=3.4in]{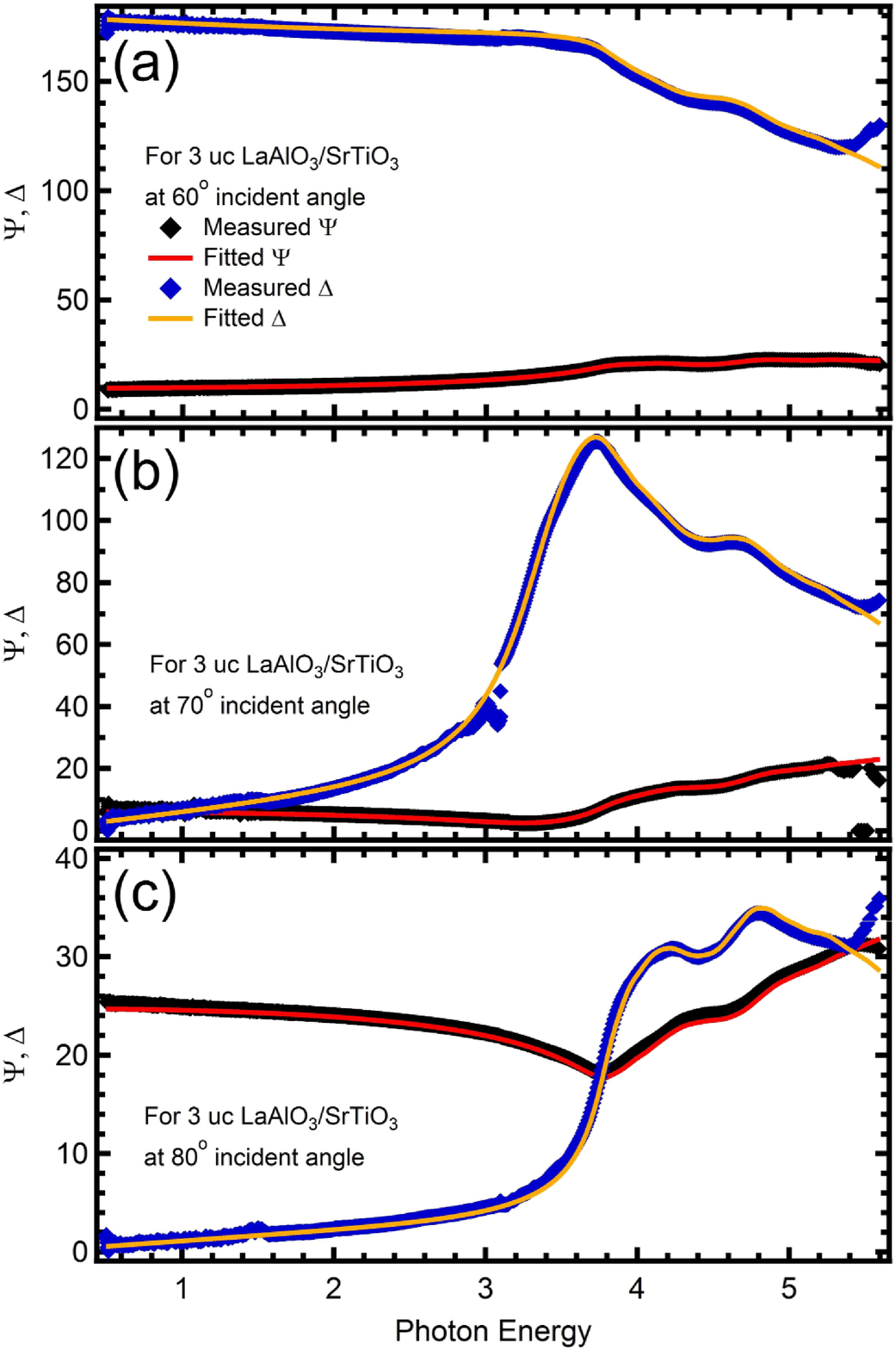}
\caption{Comparison between the experimentally-measured $\Psi$ and $\Delta$ of 3 uc LaAlO$_{3}$/SrTiO$_{3}$ and their fitted values (after iteration) for all three incident angles. (a) For 60$^{\circ}$ incident angle. (b) For 70$^{\circ}$ incident angle. (c) For 80$^{\circ}$ incident angle. The fitted values match the measured $\Psi$ and $\Delta$ very well for all three incident angles, confirming the stability of the iteration. The raw data are reproduced with permission from T. C. Asmara {\it et al.}, Nat. Commun. {\bf 5}, 3663 (2014) \cite{AsmaraNatComm}. Copyright 2014 by Nature Publishing Groups (NPG).}
\label{fig:fig8}
\end{figure}

For the insulating samples (2 and 3 uc LaAlO$_{3}$/SrTiO$_{3}$), the iteration-based analysis is also performed similarly. The fitted $\Psi$ and fitted $\Delta$ after convergence that match the measured $\Psi$ and measured $\Delta$ of 2 and 3 uc LaAlO$_{3}$/SrTiO$_{3}$ are shown in Figures~\ref{fig:fig7} and \ref{fig:fig8}, respectively. For the sake of consistency and to make layer-by-layer comparison between insulating and conducting LaAlO$_{3}$/SrTiO$_{3}$ more readily apparent, the interface layer is still initially retained in the iteration process. However, as shown later in Figure~\ref{fig:fig14}, after analyzing the normalized reflectivity in the full range of 0.5 - 35 eV, the $\varepsilon(\omega)$ of the (artificial) interface layer is found to be very similar to that of bulk SrTiO$_{3}$, making insulating LaAlO$_{3}$/SrTiO$_{3}$ effectively a two-layer structure instead. This can be explained by the absence of the conducting interface in insulating LaAlO$_{3}$/SrTiO$_{3}$.

\section{Thickness-dependent iteration procedure}

\begin{figure}
\includegraphics[width=3.4in]{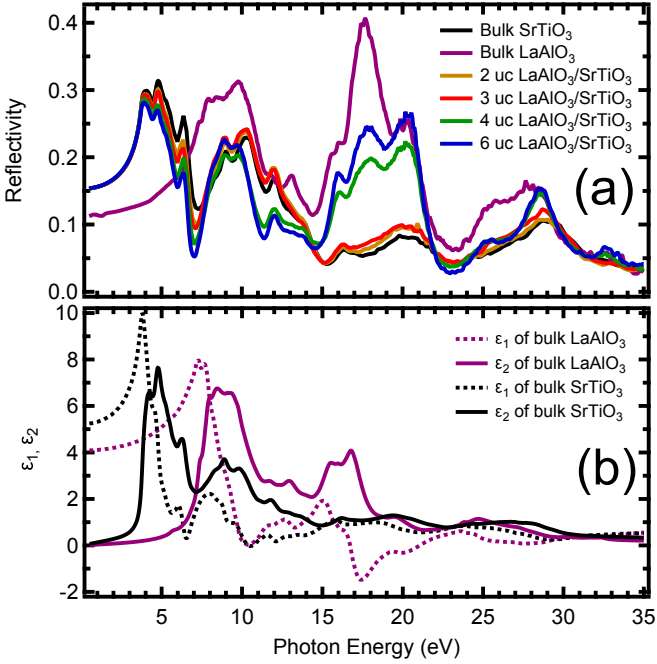}
\caption{(a) Reflectivity of LaAlO$_{3}$/SrTiO$_{3}$ at different thicknesses of LaAlO$_{3}$ film as compared to bulk LaAlO$_{3}$ and bulk SrTiO$_{3}$. (b) Complex dielectric functions, $\varepsilon(\omega)=\varepsilon_{1}(\omega)+i\varepsilon_{2}(\omega)$, of bulk LaAlO$_{3}$ and bulk SrTiO$_{3}$. The raw data are reproduced with permission from T. C. Asmara {\it et al.}, Nat. Commun. {\bf 5}, 3663 (2014) \cite{AsmaraNatComm}. Copyright 2014 by Nature Publishing Groups (NPG).}
\label{fig:fig9}
\end{figure}

From the iteration-based analysis of SE data, the $\varepsilon(\omega)$ of each individual constituent layer of LaAlO$_{3}$/SrTiO$_{3}$, along with their thicknesses, can be extracted. These quantities can then be converted into reflectivity within the 0.5 - 5.6 eV range using Eqs.~\ref{eq:eq2} - \ref{eq:eq4}, \ref{eq:eq17}, and \ref{eq:eq19}. From here, the normalization procedure of the UV-VUV reflectivity data (3.7 - 35 eV) is similar to that of bulk materials: using the SE-derived reflectivity to normalize the low-energy side (3.7 - 5.6 eV) and the off-resonance scattering considerations according to Eqs.~\ref{eq:eq5} and~\ref{eq:eq6} to normalize the high-energy side ($>$ 30 eV). The normalized reflectivity of the LaAlO$_{3}$/SrTiO$_{3}$ samples along with that of bulk LaAlO$_{3}$ and bulk SrTiO$_{3}$ is shown in Figure~\ref{fig:fig9} (a).

The same challenge present in the analysis of SE data of LaAlO$_{3}$/SrTiO$_{3}$ due to its multilayered structure is also present in analyzing the high-energy reflectivity data. Even though $d_{int}$ is already known to be $\sim$5.3 nm from the SE angle-dependent iteration analysis, it still leaves us with two unknowns (high photon energy $\varepsilon_{fLAO}$ and high photon energy $\varepsilon_{int}$), but only one equation (Eq.~\ref{eq:eq19}), which prevents a straight-forward mathematical solution. Furthermore, due to a fixed incident angle of 17.5$^{\circ}$ from the sample normal, similar angle-dependent iteration as the one done for the SE data cannot be performed. To address this, we note that Eq.~\ref{eq:eq19} can also be diversified through Eq.~\ref{eq:eq17} by varying the layer thickness \cite{SoJOSA1972}, in particular $d_{fLAO}$. It is for this reason that we have intentionally fabricated a pair of insulating samples (2 and 3 uc of LaAlO$_{3}$) and a pair of conducting samples (4 and 6 uc of LaAlO$_{3}$). Each pair has similar respective physics with only difference in $d_{fLAO}$, which can be rectified by appropriately adjusting Eq.~\ref{eq:eq17}. This means for each case (insulating and conducting), there are two unknowns and two equations for $R_{amb,multi}$, so a self-consistent iteration can be used to extract $\varepsilon(\omega)$ of each individual layer.

\begin{figure}
\includegraphics[width=3.4in]{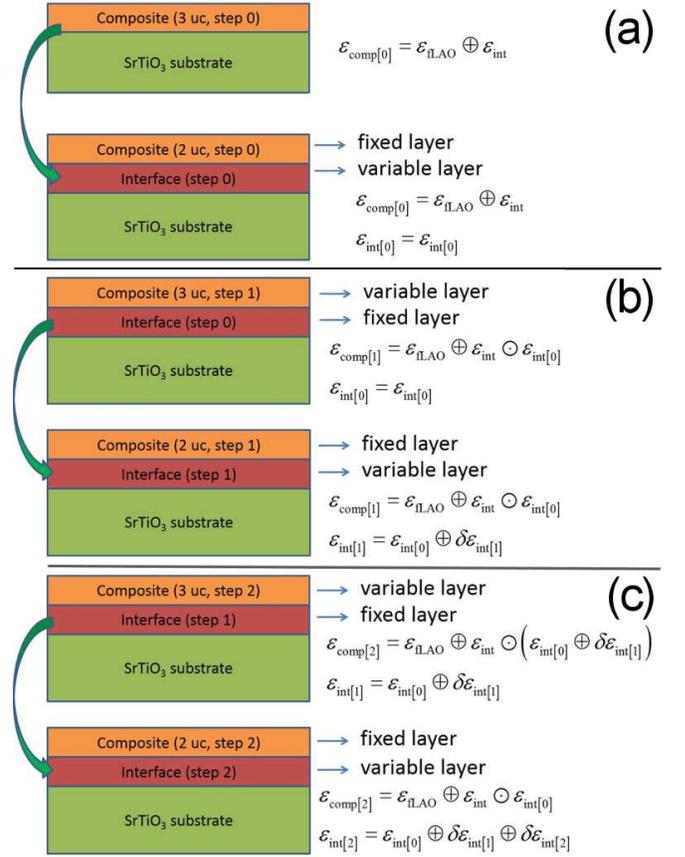}
\caption{Diagram depicting the iteration process of the 2 and 3 uc LaAlO$_{3}$/SrTiO$_{3}$ samples at (a) the initial step (step 0), (b) step 1, and (c) step 2.}
\label{fig:fig10}
\end{figure}

Let us first discuss the iteration procedure for the insulating samples (2 and 3 uc of LaAlO$_{3}$/SrTiO$_{3}$). As the initial step (step 0), the 3 uc LaAlO$_{3}$/SrTiO$_{3}$ is regarded as only having two layers: the bulk SrTiO$_{3}$ substrate below and a composite layer ($comp$) on top (Figure~\ref{fig:fig10} (a)). This composite layer represents the mixture between the unknown LaAlO$_{3}$ film and the unknown interface layer,
\begin{equation}\label{eq:eq28}
    \varepsilon_{comp[0]}=\varepsilon_{fLAO}\oplus\varepsilon_{int}.
\end{equation}
Similar to Eq.~\ref{eq:eq23}, Eq.~\ref{eq:eq28} can be explicitly expressed to the first order as,
\begin{equation}\label{eq:eq29}
    \varepsilon_{comp[0]}=\varepsilon_{fLAO} + \varepsilon_{int} \frac {\partial R_{amb,multi} / \partial \varepsilon_{int}} {\partial R_{amb,multi} / \partial \varepsilon_{fLAO}}.
\end{equation}

Because the separately-measured $\varepsilon_{STO}$ is known (Figure~\ref{fig:fig9} (b)), $\varepsilon_{comp[0]}$ can be obtained by fitting Eq.~\ref{eq:eq16} into reflectivity of 3 uc LaAlO$_{3}$/SrTiO$_{3}$ ($R_{3uc}$) using Eq.~\ref{eq:eq15} by appropriately adjusting the Drude-Lorentz oscillators that make up $\varepsilon_{comp[0]}$ (similar to the procedure used in the angle-dependent iteration, see Sec. V). After that, for the 2 uc LaAlO$_{3}$/SrTiO$_{3}$, the interface layer is added between the SrTiO$_{3}$ substrate and the composite layer so that it now is regarded as having three layers (Figure~\ref{fig:fig10} (a)). This interface layer is added for the sake of consistency and to make layer-by-layer comparison between insulating and conducting LaAlO$_{3}$/SrTiO$_{3}$ more readily apparent. The newly-fitted $\varepsilon_{comp[0]}$ along with the already-known $\varepsilon_{STO}$ is now taken as the input ($i.e.$ the composite layer becomes the "fixed layer"), and then Eq.~\ref{eq:eq19} is fitted into reflectivity of 2 uc LaAlO$_{3}$/SrTiO$_{3}$ ($R_{2uc}$) using Eq.~\ref{eq:eq15} to extract the newly-added interface layer (the "variable layer") $\varepsilon_{int[0]}$.

For the next step (step 1), the 3 uc sample is also regarded as having three layers (Figure~\ref{fig:fig10} (b)), and this time the interface layer is designated to be the input fixed layer in order to extract $\varepsilon(\omega)$ of the now-variable composite layer from $R_{3uc}$. After the extraction, the $\varepsilon_{comp}$ for this step 1 becomes
\begin{equation}\label{eq:eq30}
    \varepsilon_{comp[1]}=\varepsilon_{fLAO}\oplus\varepsilon_{int}\ominus\varepsilon_{int[0]}.
\end{equation}
Then, the focus is again shifted to $R_{2uc}$. The newly-adjusted composite layer ($\varepsilon_{comp[1]}$) is designated as the input fixed layer to extract $\varepsilon(\omega)$ of the variable interface layer,
\begin{equation}\label{eq:eq31}
   \varepsilon_{int[1]}=\varepsilon_{int[0]}\oplus\delta\varepsilon_{int[1]},
\end{equation}
from $R_{2uc}$. This $\varepsilon_{int[1]}$ is slightly different than $\varepsilon_{int[0]}$, by an amount of $\delta\varepsilon_{int[1]}$. In step 2 (Figure~\ref{fig:fig10} (c)), this procedure is repeated again, and by the end of that step $\varepsilon(\omega)$ of the layers becomes
\begin{equation}\label{eq:eq32}
    \varepsilon_{comp[2]}=\varepsilon_{fLAO}\oplus\varepsilon_{int}\ominus(\varepsilon_{int[0]}\oplus\delta\varepsilon_{int[1]})
\end{equation}
and
\begin{equation}\label{eq:eq33}
   \varepsilon_{int[2]}=\varepsilon_{int[0]}\oplus\delta\varepsilon_{int[1]}\oplus\delta\varepsilon_{int[2]}.
\end{equation}

\begin{figure}
\includegraphics[width=3.4in]{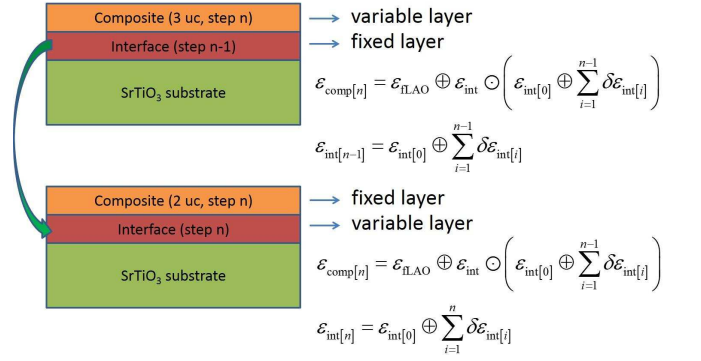}
\caption{Diagram depicting the iteration process of the 2 and 3 uc LaAlO$_{3}$/SrTiO$_{3}$ samples at a certain general step $n$.}
\label{fig:fig11}
\end{figure}

The iteration procedure is repeated until it eventually achieves convergence (see discussion below), and at a certain general step $n$ (Figure~\ref{fig:fig11}), $\varepsilon(\omega)$ of the composite and interface layers can be expressed as,
\begin{equation}\label{eq:eq34}
    \varepsilon_{comp[n]}=\varepsilon_{fLAO}\oplus\varepsilon_{int}\ominus(\varepsilon_{int[0]}\oplus\sum_{i=1}^{n-1}\delta\varepsilon_{int[i]})
\end{equation}
and
\begin{equation}\label{eq:eq35}
   \varepsilon_{int[n]}=\varepsilon_{int[0]}\oplus\sum_{i=1}^{n}\delta\varepsilon_{int[i]}.
\end{equation}

Eventually, at a certain step $N$, $\varepsilon(\omega)$ of the layers become such that,
\begin{equation}\label{eq:eq36}
   \delta\varepsilon_{int[N]}\rightarrow0,
\end{equation}
\begin{equation}\label{eq:eq37}
   \varepsilon_{int[N]}\approx\varepsilon_{int[N-1]},
\end{equation}
and
\begin{equation}\label{eq:eq38}
   \varepsilon_{comp[N]}\approx\varepsilon_{comp[N-1]}.
\end{equation}
This is our point of convergence. At this point, $\varepsilon_{int[N]}$ converges to $\varepsilon_{int}$,
\begin{equation}\label{eq:eq39}
   \varepsilon_{int[N]}=\varepsilon_{int[0]}\oplus\sum_{i=1}^{N}\delta\varepsilon_{int[i]}\rightarrow\varepsilon_{int},
\end{equation}
and $\varepsilon_{comp[N+1]}$ converges to $\varepsilon_{fLAO}$,
\begin{equation}\label{eq:eq40}
\begin{split}
   \varepsilon_{comp[N+1]}&=\varepsilon_{fLAO}\oplus\varepsilon_{int}\ominus(\varepsilon_{int[0]}\oplus\sum_{i=1}^{N}\delta\varepsilon_{int[i]}) \\
                          &\rightarrow\varepsilon_{fLAO}\oplus\varepsilon_{int}\ominus\varepsilon_{int} \\
                          &\rightarrow\varepsilon_{fLAO}.
\end{split}
\end{equation}
In other words, when this point is reached, the $\varepsilon(\omega)$ of each individual layer is able to be isolated separately, and the iteration procedure converges successfully.

\begin{figure*}
\includegraphics[width=7in]{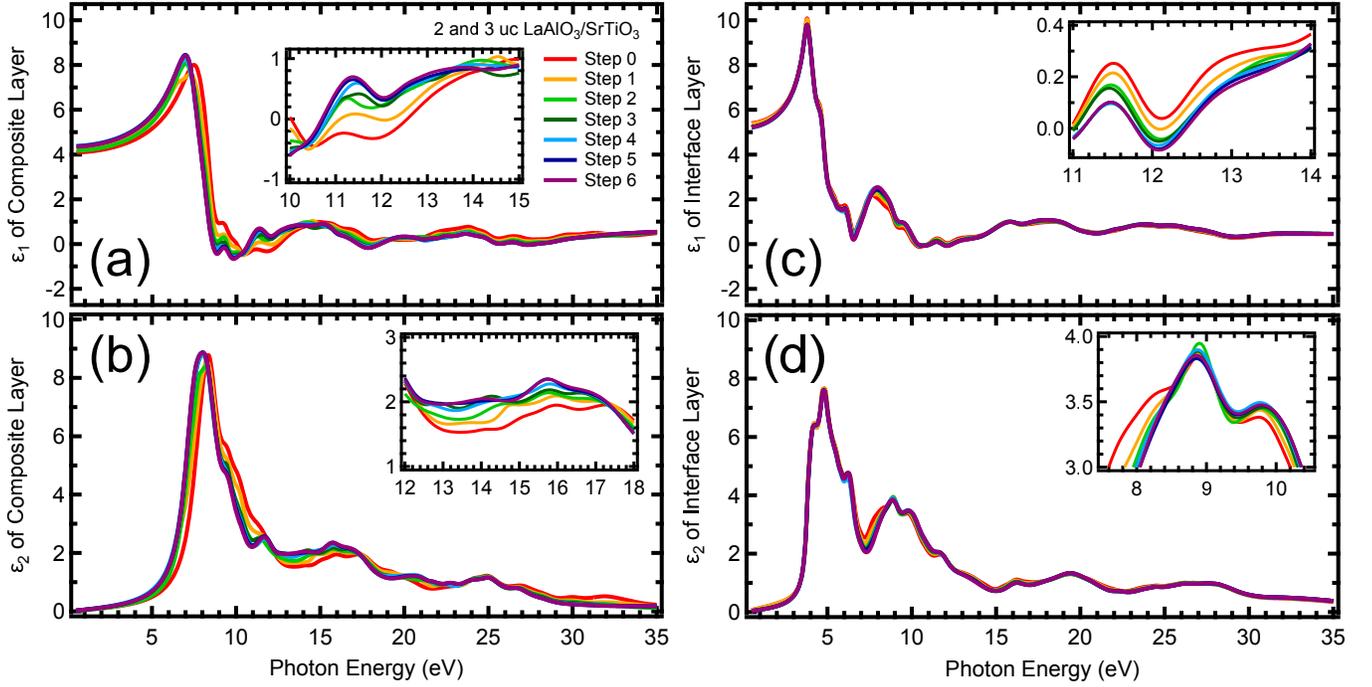}
\caption{Complex dielectric functions, $\varepsilon(\omega)=\varepsilon_{1}(\omega)+i\varepsilon_{2}(\omega)$, of composite and interface layers of the 2 and 3 uc LaAlO$_{3}$/SrTiO$_{3}$ at various steps as the thickness-dependent iteration progresses. (a) The real part of the dielectric function, $\varepsilon_{1}(\omega)$, of the composite layer. (b) The imaginary part of the dielectric function, $\varepsilon_{2}(\omega)$, of the composite layer. (c) The $\varepsilon_{1}(\omega)$ of the interface layer. (d) The $\varepsilon_{2}(\omega)$ of the interface layer. Insets show parts of the plots zoomed at various energy ranges to emphasize the evolution of $\varepsilon(\omega)$ as the iteration progresses.}
\label{fig:fig12}
\end{figure*}

In order to give a better presentation of how this iteration procedure is applied to the LaAlO$_{3}$/SrTiO$_{3}$ case study, $\varepsilon(\omega)$ of the composite and interface layers of 2 and 3 uc LaAlO$_{3}$/SrTiO$_{3}$ at various iteration steps is shown in Figure~\ref{fig:fig12}. From there, it can be seen that as the iteration progresses, the difference between $\varepsilon(\omega)$ of each consecutive step becomes progressively smaller. Eventually, the $\varepsilon(\omega)$ at step 6 becomes virtually indistinguishable to that of step 5, which means that at step 6 the iteration converges. The $\varepsilon(\omega)$ of the interface layer is successfully separated from the composite layer, and the $\varepsilon(\omega)$ of the composite layer becomes equal to the $\varepsilon(\omega)$ of the LaAlO$_{3}$ film layer.

\begin{figure}
\includegraphics[width=3.4in]{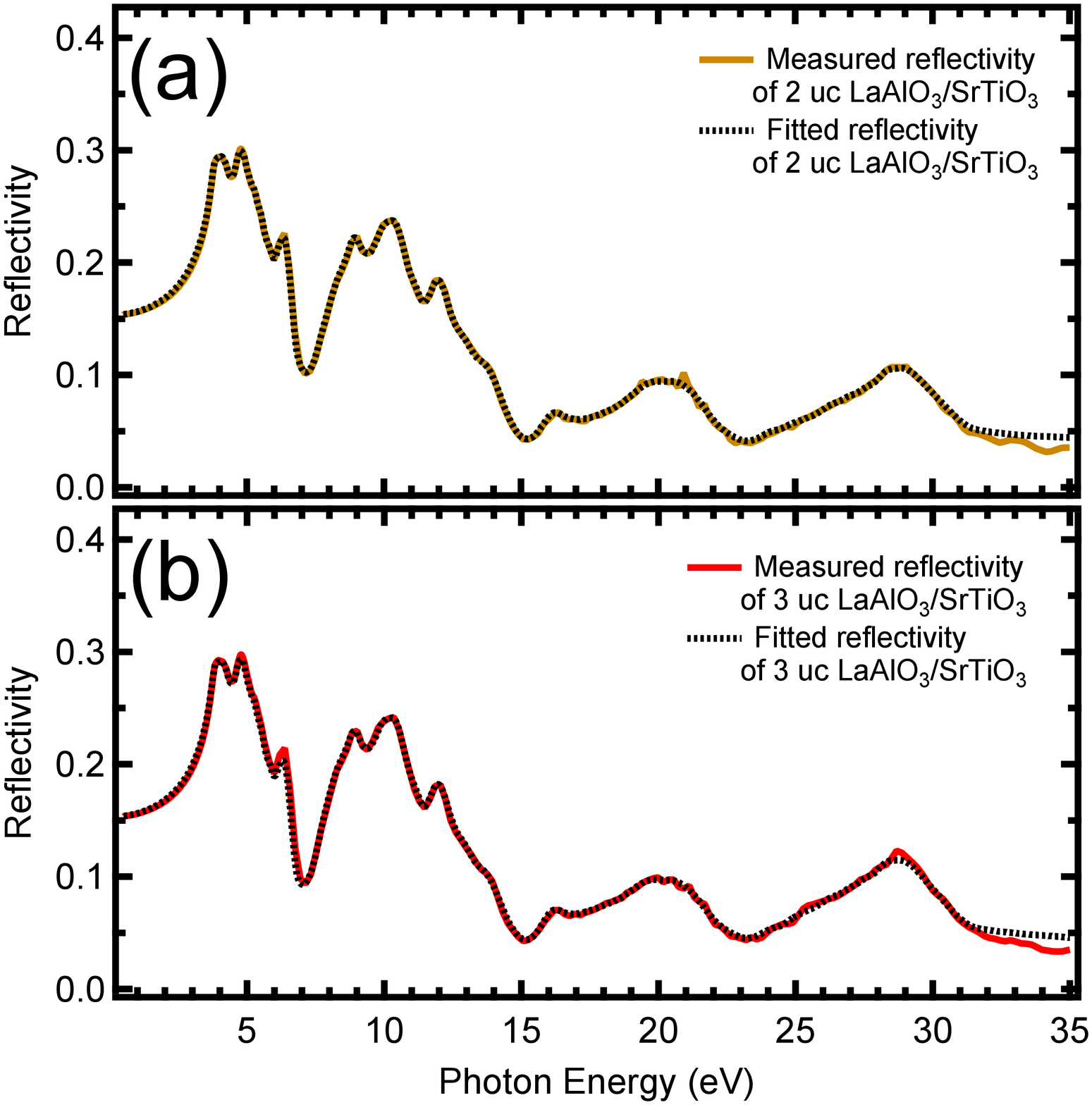}
\caption{Fitted reflectivity of insulating LaAlO$_{3}$/SrTiO$_{3}$ as compared to their experimentally-measured values after the thickness-dependent iteration convergence. (a) Fitted and measured reflectivity of 2 uc LaAlO$_{3}$/SrTiO$_{3}$. (b) Fitted and measured reflectivity of 3 uc LaAlO$_{3}$/SrTiO$_{3}$. The raw data are reproduced with permission from T. C. Asmara {\it et al.}, Nat. Commun. {\bf 5}, 3663 (2014) \cite{AsmaraNatComm}. Copyright 2014 by Nature Publishing Groups (NPG).}
\label{fig:fig13}
\end{figure}

Moreover, to further ensure the validity of the resulting $\varepsilon_{int}$ and $\varepsilon_{fLAO}$, a consistency check can be performed by inserting $\varepsilon_{int[N]}$ and $\varepsilon_{comp[N+1]}$ along with $\varepsilon_{STO}$ into Eq.~\ref{eq:eq19} via Eqs.~\ref{eq:eq2} - \ref{eq:eq4}, \ref{eq:eq17}, and \ref{eq:eq19} and confirming that the resulting $R_{amb,multi}$ can indeed reproduce the experimentally-measured reflectivity of both 2 and 3 uc LaAlO$_{3}$/SrTiO$_{3}$ simultaneously by appropriately adjusting the LaAlO$_{3}$ thickness factor in Eq.~\ref{eq:eq17} (see Figure~\ref{fig:fig13}). Thus, at the point of convergence the extracted $\varepsilon(\omega)$ of each individual layer is able to universally fit the reflectivity of both 2 and 3 uc LaAlO$_{3}$/SrTiO$_{3}$.

For the 4 and 6 uc LaAlO$_{3}$/SrTiO$_{3}$ samples (the conducting case), the iteration process should essentially be the same as the iteration procedure for the insulating case. However, in practice it can be more complex than that. This is because in the conducting sample the interface layer is conducting, so its optical properties can be very different than that of bulk SrTiO$_{3}$. This makes the two-layered structure used in the first part of step 0 unsuitable to model the strictly three-layered system. To circumvent it, this first part of step 0 can be skipped entirely. Instead, the 4 uc LaAlO$_{3}$/SrTiO$_{3}$ is directly regarded from the start as having three layers: the composite layer on top, the interface layer in the middle, and the bulk SrTiO$_{3}$ substrate at the bottom. The initial guess of $\varepsilon(\omega)$ of the composite layer can be set as equal to $\varepsilon(\omega)$ of bulk LaAlO$_{3}$, $\varepsilon_{comp[0]}=\varepsilon_{bLAO}$, and from here the iteration can be continued as normal until convergence is achieved and the actual $\varepsilon_{2}(\omega)$ of both the interface and the LaAlO$_{3}$ film layers are found.

\begin{figure*}
\includegraphics[width=7in]{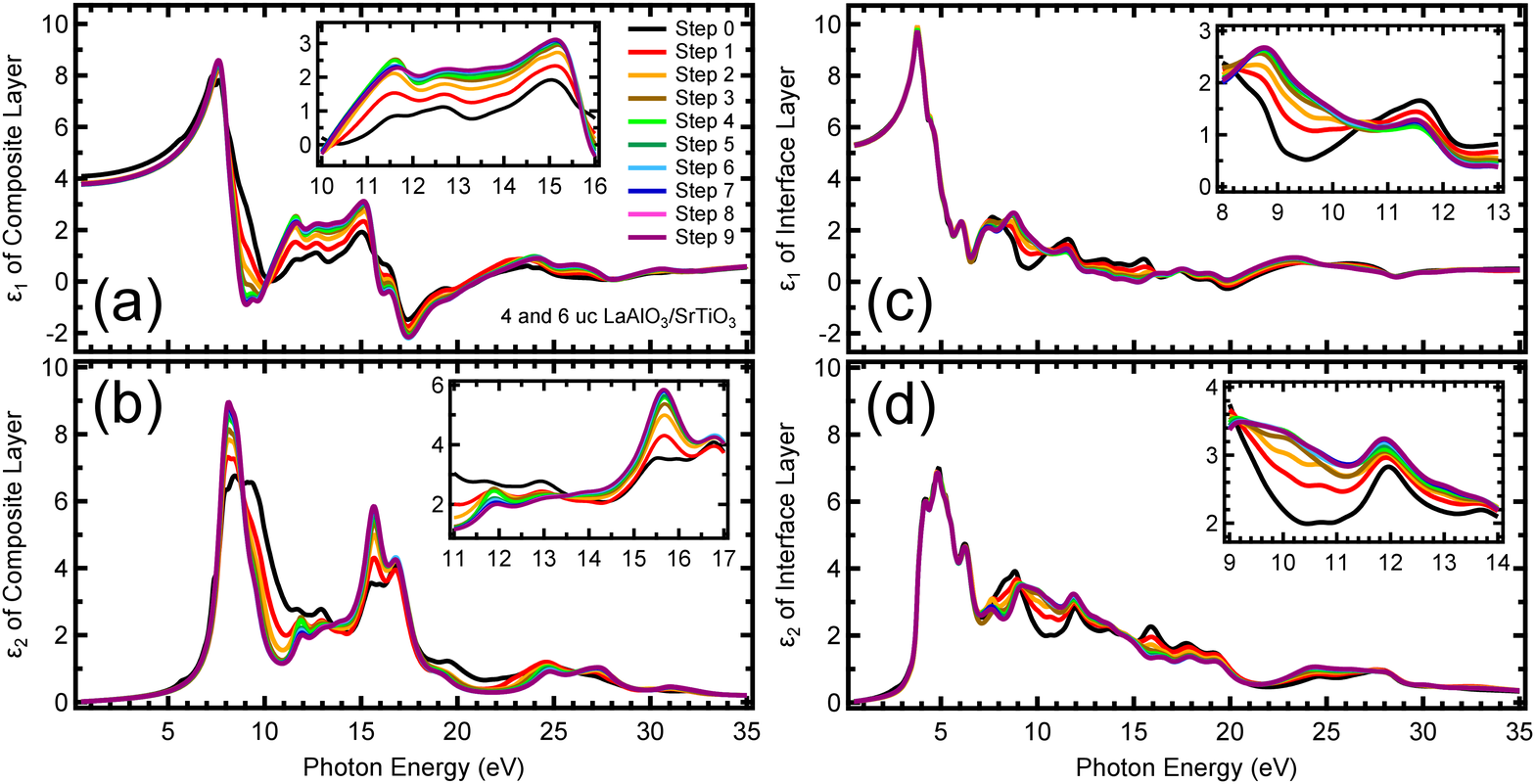}
\caption{Complex dielectric functions, $\varepsilon(\omega)=\varepsilon_{1}(\omega)+i\varepsilon_{2}(\omega)$, of composite and interface layers of the 4 and 6 uc LaAlO$_{3}$/SrTiO$_{3}$ at various steps as the thickness-dependent iteration using the modified initial guess progresses. (a) The real part of the dielectric function, $\varepsilon_{1}(\omega)$, of the composite layer. (b) The imaginary part of the dielectric function, $\varepsilon_{2}(\omega)$, of the composite layer. (c) The $\varepsilon_{1}(\omega)$ of the interface layer. (d) The $\varepsilon_{2}(\omega)$ of the interface layer. Insets show parts of the plots zoomed at various energy ranges to emphasize the evolution of $\varepsilon(\omega)$ as the iteration progresses.}
\label{fig:fig14}
\end{figure*}

\begin{figure}
\includegraphics[width=3.4in]{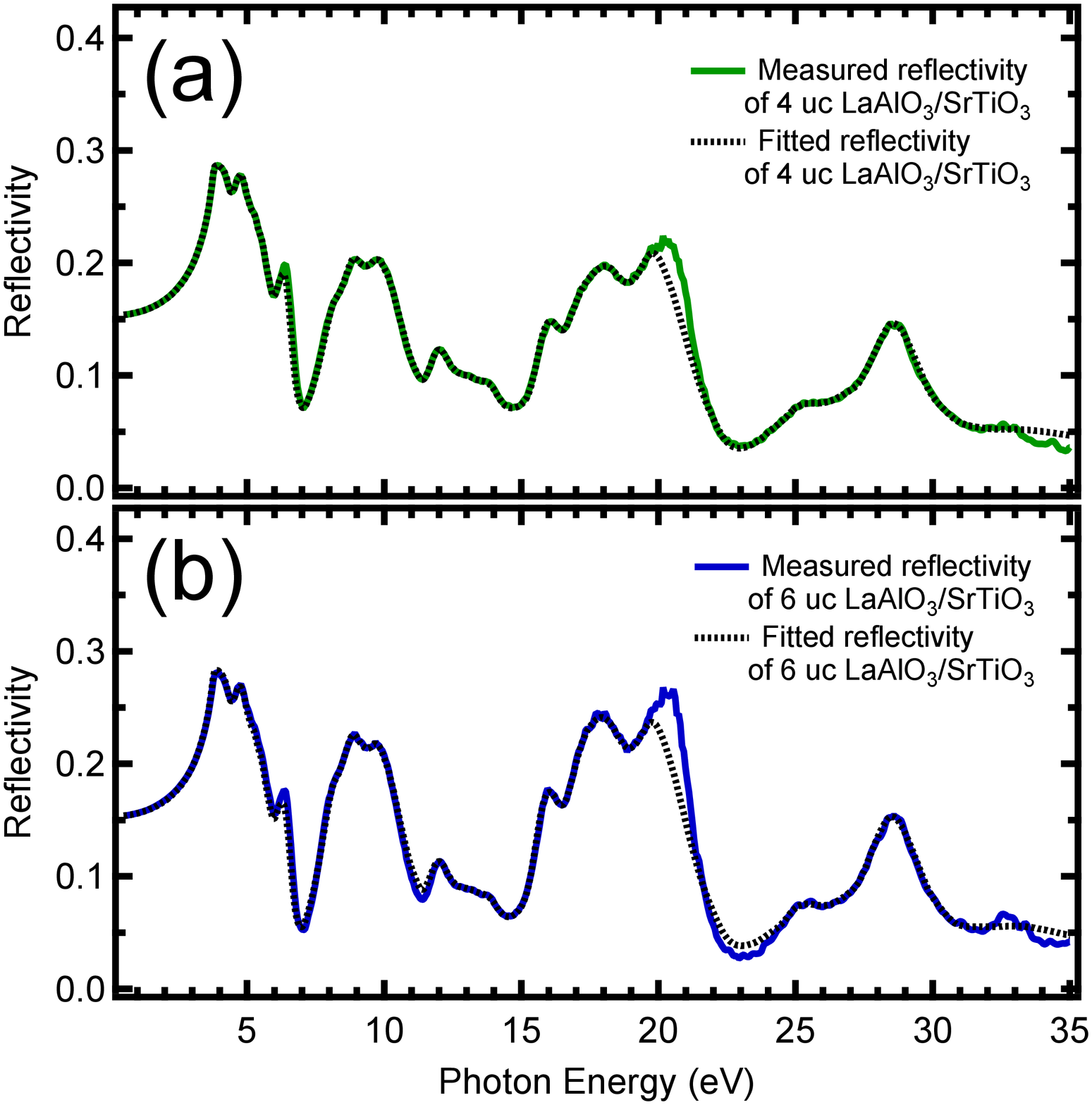}
\caption{Fitted reflectivity of conducting LaAlO$_{3}$/SrTiO$_{3}$ as compared to their experimentally-measured values after the thickness-dependent iteration convergence. (a) Fitted and measured reflectivity of 4 uc LaAlO$_{3}$/SrTiO$_{3}$. (b) Fitted and measured reflectivity of 6 uc LaAlO$_{3}$/SrTiO$_{3}$. The raw data are reproduced with permission from T. C. Asmara {\it et al.}, Nat. Commun. {\bf 5}, 3663 (2014) \cite{AsmaraNatComm}. Copyright 2014 by Nature Publishing Groups (NPG).}
\label{fig:fig15}
\end{figure}

Figure~\ref{fig:fig14} shows the $\varepsilon(\omega)$ of the composite and interface layers of 4 and 6 uc LaAlO$_{3}$/SrTiO$_{3}$ at various iteration steps. Similar with the 2 and 3 uc case, as the iteration progresses the difference between $\varepsilon(\omega)$ of each consecutive step becomes progressively smaller, and eventually at step 9 the iteration converges and $\varepsilon(\omega)$ of the LaAlO$_{3}$ film and the interface layer are successfully separated.

For consistency check, the fitted reflectivity of the 4 and 6 uc LaAlO$_{3}$/SrTiO$_{3}$ after convergence is compared to their measured values in Figure~\ref{fig:fig15}. It shows that the resulting $\varepsilon(\omega)$ of LaAlO$_{3}$ film and interface layer are indeed able to closely reproduce the experimentally-measured reflectivity of both 4 and 6 uc LaAlO$_{3}$/SrTiO$_{3}$ simultaneously, by appropriately adjusting the LaAlO$_{3}$ thickness factor in Eq.~\ref{eq:eq17}.

\section{Results: complex dielectric functions of LAO/STO}

\begin{figure*}
\includegraphics[width=7in]{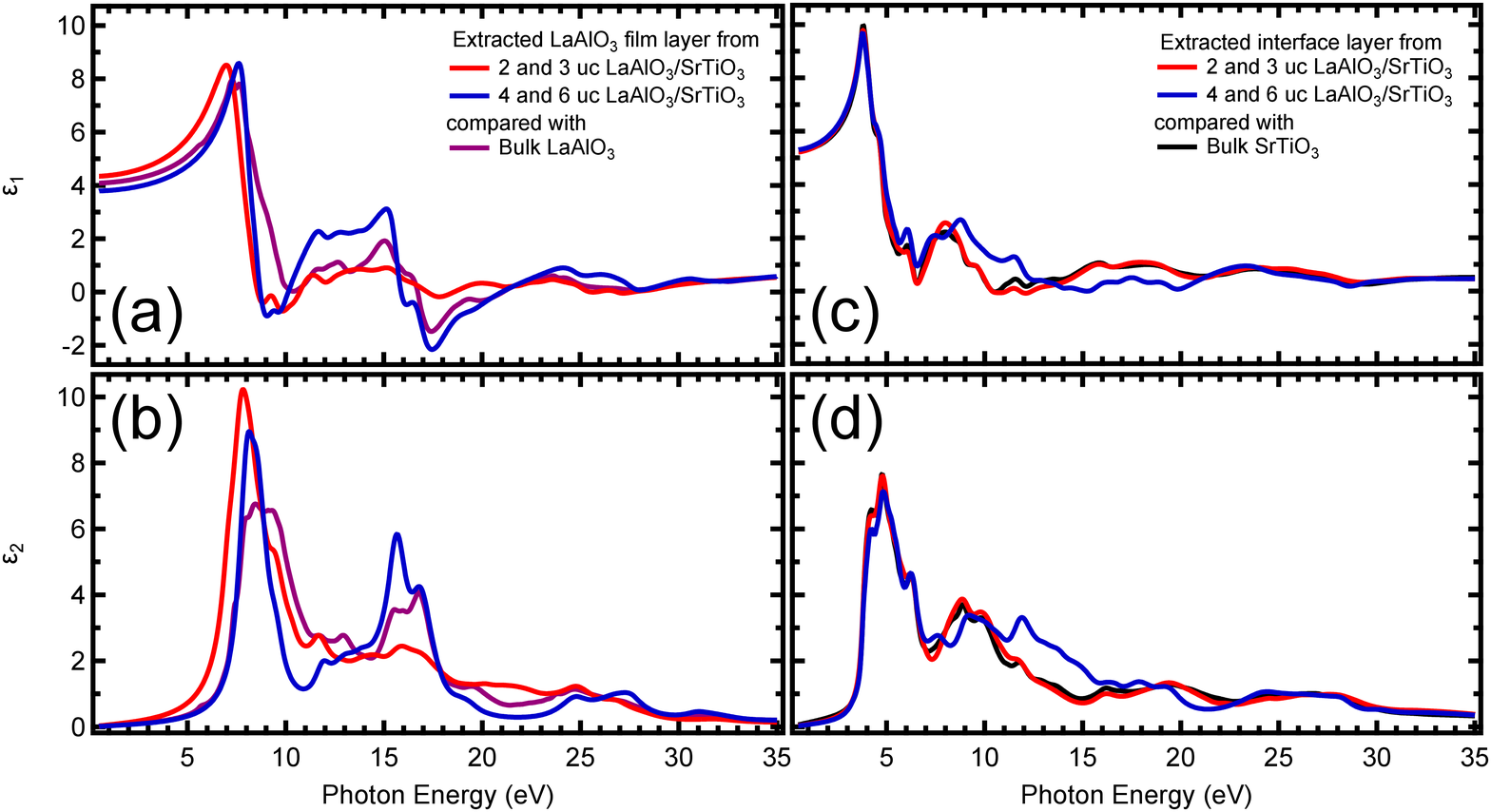}
\caption{Complex dielectric functions, $\varepsilon(\omega)=\varepsilon_{1}(\omega)+i\varepsilon_{2}(\omega)$, of each layer of 2, 3, 4, and 6 uc LaAlO$_{3}$/SrTiO$_{3}$, extracted from reflectivity using the self-consistent iteration procedure. (a) The real part, $\varepsilon_{1}(\omega)$, of dielectric function of LaAlO$_{3}$ film, as compared to bulk LaAlO$_{3}$3. (b) The imaginary part, $\varepsilon_{2}(\omega)$, of dielectric function of LaAlO$_{3}$ film, as compared to bulk LaAlO$_{3}$. (c) The $\varepsilon_{1}(\omega)$ of the interface layer, as compared to bulk SrTiO$_{3}$. (d) The $\varepsilon_{2}(\omega)$ of the interface layer, as compared to bulk SrTiO$_{3}$. Note that the $\varepsilon(\omega)$ of the LaAlO$_{3}$ film of the 2 and 3 uc LaAlO$_{3}$/SrTiO$_{3}$ are identical due to the nature of the thickness-dependent iteration process. The same is true for the $\varepsilon(\omega)$ of the interface layer and for the 4 and 6 uc LaAlO$_{3}$/SrTiO$_{3}$ case. The raw data are reproduced from Asmara {\it et al.} \cite{AsmaraNatComm}.}
\label{fig:fig16}
\end{figure*}

Figure~\ref{fig:fig16} summarizes the analysis results of LaAlO$_{3}$-thickness-dependent LaAlO$_{3}$/SrTiO$_{3}$ using the self-consistent iteration procedure. It shows the extracted $\varepsilon(\omega)$ of LaAlO$_{3}$ film and interface layer for both the insulating (2 and 3 uc LaAlO$_{3}$/SrTiO$_{3}$) and conducting (4 and 6 uc LaAlO$_{3}$/SrTiO$_{3}$) cases as compared to bulk LaAlO$_{3}$ and bulk SrTiO$_{3}$. For LaAlO$_{3}$ film, it can be seen (Figures~\ref{fig:fig16} (a) and (b)) that the $\varepsilon(\omega)$ of LaAlO$_{3}$ film for both the insulating and conducting LaAlO$_{3}$/SrTiO$_{3}$ are very different than bulk LaAlO$_{3}$ and also as compared to each other. This indicates that there are significant differences in band structure and orbital occupancy among the different forms (bulk or film) and environments (insulating LaAlO$_{3}$/SrTiO$_{3}$ or conducting LaAlO$_{3}$/SrTiO$_{3}$) of LaAlO$_{3}$, and careful investigation of these differences might play a role in revealing the mechanisms behind the plethora of interesting phenomena of LaAlO$_{3}$/SrTiO$_{3}$.

For the interface layer, Figures~\ref{fig:fig16} (c) and (d) show that $\varepsilon(\omega)$ at the interface of insulating samples (2 and 3 uc LaAlO$_{3}$/SrTiO$_{3}$) is very similar to that of bulk SrTiO$_{3}$, which makes insulating LaAlO$_{3}$/SrTiO$_{3}$ effectively a two-layer system. On the other hand, interestingly for conducting samples (4 and 6 uc LaAlO$_{3}$/SrTiO$_{3}$) there is a new feature around 8 - 12 eV for $\varepsilon_{1}(\omega)$ and 11 - 16 eV for $\varepsilon_{2}(\omega)$, which can be a characteristic of the 2DEG that emerges as the LaAlO$_{3}$/SrTiO$_{3}$ interface becomes conducting. Apart from this new feature, the $\varepsilon(\omega)$ of the interface layer of conducting LaAlO$_{3}$/SrTiO$_{3}$ quite closely resembles the $\varepsilon(\omega)$ of bulk SrTiO$_{3}$, which indicates that the interface layer is SrTiO$_{3}$-like, and that the conducting layer mostly resides in SrTiO$_{3}$ side rather than LaAlO$_{3}$.

To investigate the $\varepsilon(\omega)$ of each layer of LaAlO$_{3}$/SrTiO$_{3}$ more thoroughly, knowledge of the band structures of LaAlO$_{3}$ and SrTiO$_{3}$ is needed to identify the optical transition of each peak in the $\varepsilon(\omega)$ spectra. From this, information about the relative changes in orbital occupancy and charge transfers among the energy bands that participate in the optical transitions can be obtained. However, such detailed discussion about the interesting physics of LaAlO$_{3}$/SrTiO$_{3}$ is beyond the scope of this paper, and thus is covered in our other studies \cite{AsmaraNatComm}.

\section{Errors analysis}

The mean squared errors (MSE) associated with the reflectivity fitting process involved in the iteration procedure can be estimated using \cite{Fujiwara2007}
\begin{equation}\label{eq:eq41}
   |\frac{\delta R}{R}|_{fit}^{2}=\frac{1}{M-1}\sum_{j=1}^{M}|\frac{R_{ex}(\omega_{j})-R_{fit}(\omega_{j})}{R_{ex}(\omega_{j})}|^{2},
\end{equation}
where $M$ is the number of measurement points, while $|\frac{\delta R}{R}|_{fit}$, $R_{ex}$, and $R_{fit}$ are the reflectivity fitting errors, the experimentally-measured reflectivity, and the fitted reflectivity, respectively. To obtain the corresponding fitting error for $\varepsilon$, $|\frac{\delta\varepsilon}{\varepsilon}|_{fit}$, in the first approximation the errors can be propagated using
\begin{equation}\label{eq:eq42}
   |\frac{\delta\varepsilon}{\varepsilon}|_{fit}=|1-\frac{1}{\sqrt{|\varepsilon}|}-\frac{1}{|\varepsilon|}||\frac{\delta R_{fit}}{R}|,
\end{equation}
which is based on Eqs.~\ref{eq:eq2} - \ref{eq:eq4}. The absolute fitting error of $\varepsilon$, $|\delta\varepsilon|_{fit}$ is then obtained by multiplying $|\frac{\delta\varepsilon}{\varepsilon}|_{fit}$ with $|\varepsilon|$ (note that the absolute fitting errors of other quantities are also estimated this way). For example, Table~\ref{tab:tab1} shows the $|\frac{\delta R}{R}|_{fit}$ of the thickness-dependent iteration of conducting LaAlO$_{3}$/SrTiO$_{3}$ and how it propagates into $|\delta\varepsilon|_{fit}$. The fitting errors of the thickness-dependent iteration of the insulating LaAlO$_{3}$/SrTiO$_{3}$ are similar to the values shown in Table~\ref{tab:tab1}.

\begin{table}
  \caption{Fitting errors of the thickness-dependent iteration of conducting LaAlO$_{3}$/SrTiO$_{3}$. Since $R$ and $\varepsilon$ are $\lambda$-dependent, $|\frac{\delta R}{R}|_{fit}$ is a quadratic average over all wavelength points, while $|\delta R|_{fit}$ and $|\delta\varepsilon|_{fit}$ are absolute averages over all wavelength data points.}
  \label{tab:tab1}
  \setlength{\tabcolsep}{17.5pt}
  \renewcommand{\arraystretch}{1.5}
  \begin{center}
  \begin{tabular}{ l c c }
  \hline
  \hline
  Sample & $|\frac{\delta R}{R}|_{fit}$ & $|\delta R|_{fit}$ \\
  \hline
  4 uc LaAlO$_{3}$/SrTiO$_{3}$ & 4.4\% & 0.003 \\
  6  uc LaAlO$_{3}$/SrTiO$_{3}$ & 9.8\% & 0.005 \\
  Average & 7.8\% & 0.004 \\
  \hline
  $\varepsilon$ & $|\delta\varepsilon|_{fit}$ \\
  \hline
  Re[$\varepsilon_{fLAO}$] & 0.06 \\
  Im[$\varepsilon_{fLAO}$] & 0.09 \\
  Re[$\varepsilon_{int}$]& 0.10 \\
  Im[$\varepsilon_{int}$] & 0.07 \\
  \hline
  \hline
\end{tabular}
\end{center}
\end{table}

For SE data, since the angle-dependent iteration is performed by fitting the $\Psi$ and $\Delta$ data, the corresponding $|\frac{\delta\varepsilon}{\varepsilon}|_{fit}$ can be estimated as follows. First, the MSE asociated with the fitting of $\rho$, $|\frac{\delta\rho}{\rho}|_{fit}$, can be estimated based on Eq.~\ref{eq:eq1} using
\begin{equation}\label{eq:eq43}
\begin{split}
   |\frac{\delta\rho}{\rho}|_{fit}^{2}=|\frac{\delta\tan\Psi}{\tan\Psi}|_{fit}^{2}+|\frac{\delta\Delta}{\Delta}|_{fit}^{2},
\end{split}
\end{equation}
where $|\frac{\delta\tan\Psi}{\tan\Psi}|_{fit}^{2}$ and $|\frac{\delta\Delta}{\tan\Delta}|_{fit}^{2}$ are obtained by substituting the $R$ in Eq.~\ref{eq:eq42} with $\tan\Psi$ and $\Delta$, respectively. Then, since $\rho$ is essentially a ratio of reflection coefficients, in the first approximation $|\frac{\delta\rho}{\rho}|_{fit}$ can be approximated to be the same as the fitting error for $r$, $|\frac{\delta r}{r}|_{fit}$. From Eq.~\ref{eq:eq43}, $|\frac{\delta\varepsilon}{\varepsilon}|_{fit}$ can be propagated from $|\frac{\delta r}{r}|_{fit}$ using Eq.~\ref{eq:eq42} by considering that $R=|r|^2$ and thus $|\frac{\delta R}{R}|_{fit}=2|\frac{\delta r}{r}|_{fit}$. For example, Table~\ref{tab:tab2} shows the $|\frac{\delta \rho}{\rho}|_{fit}$ of the angle-dependent iteration of 6 uc LaAlO$_{3}$/SrTiO$_{3}$ and how it propagates into $|\delta\varepsilon|_{fit}$. The fitting errors of the angle-dependent iteration of other LaAlO$_{3}$/SrTiO$_{3}$ samples are similar to the values shown in Table~\ref{tab:tab2}.

\begin{table}
  \caption{Fitting errors of the angle-dependent iteration of 6 uc LaAlO$_{3}$/SrTiO$_{3}$. Since $\Psi$, $\Delta$, $\rho$, and $\varepsilon$ are $\lambda$-dependent, $|\frac{\delta\Psi}{\Psi}|_{fit}$, $|\frac{\delta\Delta}{\Delta}|_{fit}$, and $|\frac{\delta\rho}{\rho}|_{fit}$ are quadratic averages over all wavelength points, while $|\delta\Psi|_{fit}$, $|\delta\Delta|_{fit}$, and $|\delta\varepsilon|_{fit}$ are absolute averages over all wavelength data points.}
  \label{tab:tab2}
  \setlength{\tabcolsep}{5.5pt}
  \renewcommand{\arraystretch}{1.5}
  \begin{center}
  \begin{tabular}{ l c c c c c }
  \hline
  \hline
  Angle & $|\frac{\delta\Psi}{\Psi}|_{fit}$ & $|\delta\Psi|_{fit}$ & $|\frac{\delta\Delta}{\Delta}|_{fit}$ & $|\delta\Delta|_{fit}$ & $|\frac{\delta\rho}{\rho}|_{fit}$ \\
  \hline
  60$^{\circ}$ & 1.1\% & 0.10$^{\circ}$ & 0.5\% & 0.57$^{\circ}$ & 1.2\% \\
  70$^{\circ}$ & 3.5\% & 0.16$^{\circ}$ & 6.5\% & 1.09$^{\circ}$ & 7.4\%\\
  80$^{\circ}$ & 1.1\% & 0.17$^{\circ}$ & 8.8\% & 0.29$^{\circ}$ & 8.9\%\\
  Average & 2.2\% & 0.14$^{\circ}$ & 6.3\% & 0.65$^{\circ}$ & 6.7\%\\
  \hline
  $\varepsilon$ & $|\delta\varepsilon|_{fit}$ & & & & \\
  \hline
  Re[$\varepsilon_{fLAO}$] & 0.15 & & & & \\
  Im[$\varepsilon_{fLAO}$] & 0.16 & & & & \\
  Re[$\varepsilon_{int}$]& 0.24 & & & & \\
  Im[$\varepsilon_{int}$] & 0.14 & & & & \\
  \hline
  \hline
\end{tabular}
\end{center}
\end{table}

Meanwhile, in the first approximation, the fitting error of $d_{int}$, $|\frac{\delta d_{int}}{d_{int}}|_{fit}$, can be propagated from $|\frac{\delta\rho}{\rho}|_{fit}$ using
\begin{equation}\label{eq:eq44}
\begin{split}
   |\frac{\delta d_{int}}{d_{int}}|_{fit}=&\frac{|r_{amb,LAO}|+|r_{LAO,int}|+|r_{int,STO}|}{2|r_{int,STO}|(1-|r_{amb,LAO}|^{2}-|r_{LAO,int}|^{2})}\\
                                          &\times\frac{1}{|\delta_{int}|}|\frac{\delta\rho}{\rho}|_{fit},
\end{split}
\end{equation}
 which is based on Eqs.~\ref{eq:eq1} and \ref{eq:eq19}. However, due to the thin film limit, $|\delta_{int}|$ is quite small, which leads to a very large $|\delta d_{int}|_{fit}$ of several nanometers, comparable to the obtained $d_{int}$ value of $\sim$5.3 nm. On the other hand, Figures~\ref{fig:fig5} (b) and (c) show that even when several iterations are performed with different initial guesses for $d_{int}$, they are still able to converge accurately to a $d_{int}$ value of $\sim$5.3 nm, with a small standard deviation of only $\sim$0.1 nm. Thus, this means that as long as the initial guess is within a reasonable limit from the actual value of $d_{int}$ and the iteration is able to converge successfully, the uncertainty of $d_{int}$ obtained from the iteration procedure should be much smaller than what $|\delta d_{int}|_{fit}$ propagated from $|\frac{\delta\rho}{\rho}|_{fit}$ would otherwise suggest. For this reason, a good initial guess, for example from the results of other techniques, is preferable in order to have a more accurate and efficient iteration. This also applies for the uncertainties of other quantities, such as the uncertainties of $\varepsilon_{fLAO}$ and $\varepsilon_{int}$ shown in Tables~\ref{tab:tab1} and \ref{tab:tab2}.

\section{Photon penetration depth analysis}

Photon penetration depth, $D$, of a material is defined as the depth at which the intensity of the incident light, $I$, drops to $1/e$ of its initial value, $I_{0}$, where $e$ is the natural constant. It can be obtained from the $\varepsilon(\omega)$ of the material according to \cite{Born2003}
\begin{equation}\label{eq:eq45}
   D=\frac{\lambda\sqrt{\varepsilon_{1}}}{2\pi\varepsilon_{2}}.
\end{equation}
If the material is multilayered like LaAlO$_{3}$/SrTiO$_{3}$, the intensity drop depends on the penetration depth of each constituent material, in this case the LaAlO$_{3}$ film, the interface layer, and the SrTiO$_{3}$ substrate,
\begin{equation}\label{eq:eq46}
   I(z)=I_{0}\exp[-(\frac{d_{fLAO}}{D_{fLAO}}+\frac{d_{int}}{D_{int}}+\frac{z-d_{fLAO}-d_{int}}{D_{STO}})],
\end{equation}
where $z > (d_{fLAO}+d_{int})$ is along the direction perpendicular to and measured from the surface of the heterostructure. From Eq.~\ref{eq:eq46}, the effective penetration depth, $D_{eff}$, of LaAlO$_{3}$/SrTiO$_{3}$ can thus be expressed as,
\begin{equation}\label{eq:eq47}
\begin{split}
   D_{eff}=&(1-\frac{D_{STO}}{D_{fLAO}})d_{fLAO}+(1-\frac{D_{STO}}{D_{int}})d_{int} \\
           &+D_{STO}.
\end{split}
\end{equation}

\begin{figure}
\includegraphics[width=3.4in]{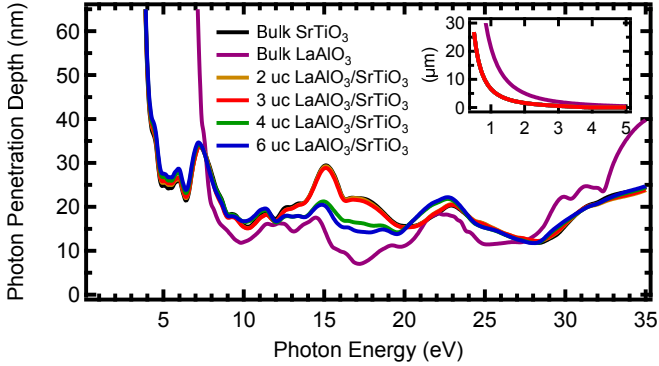}
\caption{Effective photon penetration depth of 2, 3, 4, and 6 uc LaAlO$_{3}$/SrTiO$_{3}$ at 17.5$^{\circ}$ incident angle along with that of bulk LaAlO$_{3}$ and bulk SrTiO$_{3}$.}
\label{fig:fig17}
\end{figure}

The $D_{eff}$ of the LaAlO$_{3}$/SrTiO$_{3}$ samples along with that of bulk LaAlO$_{3}$ and bulk SrTiO$_{3}$ at 17.5$^{\circ}$ incident angle is shown in Figure~\ref{fig:fig17}. From the figure, it can be seen that the $D_{eff}$ of all samples is around 10 - 40 nm above 5 eV and up to 30 $\mu$m below it, which is more than sufficient to cover the LaAlO$_{3}$ film thickness of 1 - 2 nm and the interface thickness of $\sim$ 5 nm. This means that the photon can indeed probe the interface thoroughly, and even able to penetrate deep into the SrTiO$_{3}$ substrate.

\section{Extensions and limitations of the iteration procedure}

The iteration procedure is not limited to the analysis of LaAlO$_{3}$/SrTiO$_{3}$, and it can be applied to analyze various other multilayer systems, even if those systems might have different numbers of unknown parameters than what is discussed in this paper. For example, if three unknown films, each with unknown thickness, are deposited on top of a known substrate, there will be six unknown parameters in total (three unknown dielectric functions and 3 unknown thicknesses). To perform an angle-dependent iteration on the system, the SE measurement needs to be done at six different incident angles, so that the number of equations (Eq.~\ref{eq:eq1}) can match the number of unknowns via Eq.~\ref{eq:eq17}. The incident angles should be chosen such that the corresponding differences in $\delta$ can give rise to relatively large variations in the measured $\Psi$ and $\Delta$ spectra, so that the iteration can be performed more efficiently. The iteration can then be performed by cycling through the $\Psi$ and $\Delta$ data measured at these 6 incident angles. Of course, with more unknown parameters the complexity also increases, which means more resources are needed to successfully converge the iteration process.

For the thickness-dependent iteration, assuming that the thicknesses of the layers are known from the angle-dependent iteration and/or other methods, and can be controlled during growth, the above example still leaves us with three unknown parameters, which are the $\varepsilon(\omega)$ of each of the 3 film layers. To perform the thickness-dependent iteration, we need to also prepare three samples with slightly different thicknesses of those 3 films. Again, to make the iteration procedure more efficient, the thicknesses should be chosen such that the corresponding differences in $\delta$ can give rise to relatively large variations in the measured reflectivity spectra. The iteration can then be performed by cycling through the reflectivity of those three samples, because the number of equations (equivalent of Eqs.~\ref{eq:eq16} and~\ref{eq:eq19}, extended to four layers) now matches the number of unknown parameters via Eq.~\ref{eq:eq17}.

However, despite its potentials, there are still some limitations inherent especially to the thickness-dependent iteration method. The first is sample variance. Since multiple samples are needed to perform the thickness-dependent iteration, slight differences in properties among the samples (for example due to slightly different growth conditions each time the samples are prepared) can accumulate to rather large deviations. Because of this, extra care is needed to ensure that each sample is prepared within almost identical environment. Secondly, it also needs to be ensured that any variation in reflectivity among the samples is only due to the different film thicknesses involved ($i.e.$ only due to Eq.~\ref{eq:eq17}). Film thickness differences should not significantly modified the internal properties of the samples, because otherwise it will render the thickness-dependent iteration procedure invalid. This is why the iteration cannot be performed between the insulating 3 uc LaAlO$_{3}$/SrTiO$_{3}$ and the conducting 4 uc LaAlO$_{3}$/SrTiO$_{3}$, since their inherent properties are modified by the increase of the LaAlO$_{3}$ thickness. Furthermore, due to this requirement, the resulting $\varepsilon(\omega)$ of each layer is identical for all the samples involved in a particular thickness-dependent iteration. For instance, $\varepsilon(\omega)$ of the LaAlO$_{3}$ film of the 2 and 3 uc LaAlO$_{3}$/SrTiO$_{3}$ in Figure~\ref{fig:fig16} are identical to each other. The same is true for the $\varepsilon(\omega)$ of the interface layer and for the 4 and 6 uc LaAlO$_{3}$/SrTiO$_{3}$ case.

These two limitations are not applicable to the angle-dependent iteration, because the measurements at the different incident angles are still performed on the same sample each time. This eliminates the concern about sample variance, and because the iteration is performed on data measured from only one sample, the converged results are also unique to that particular sample.

\section{Conclusions}

\begin{figure*}
\includegraphics[width=7in]{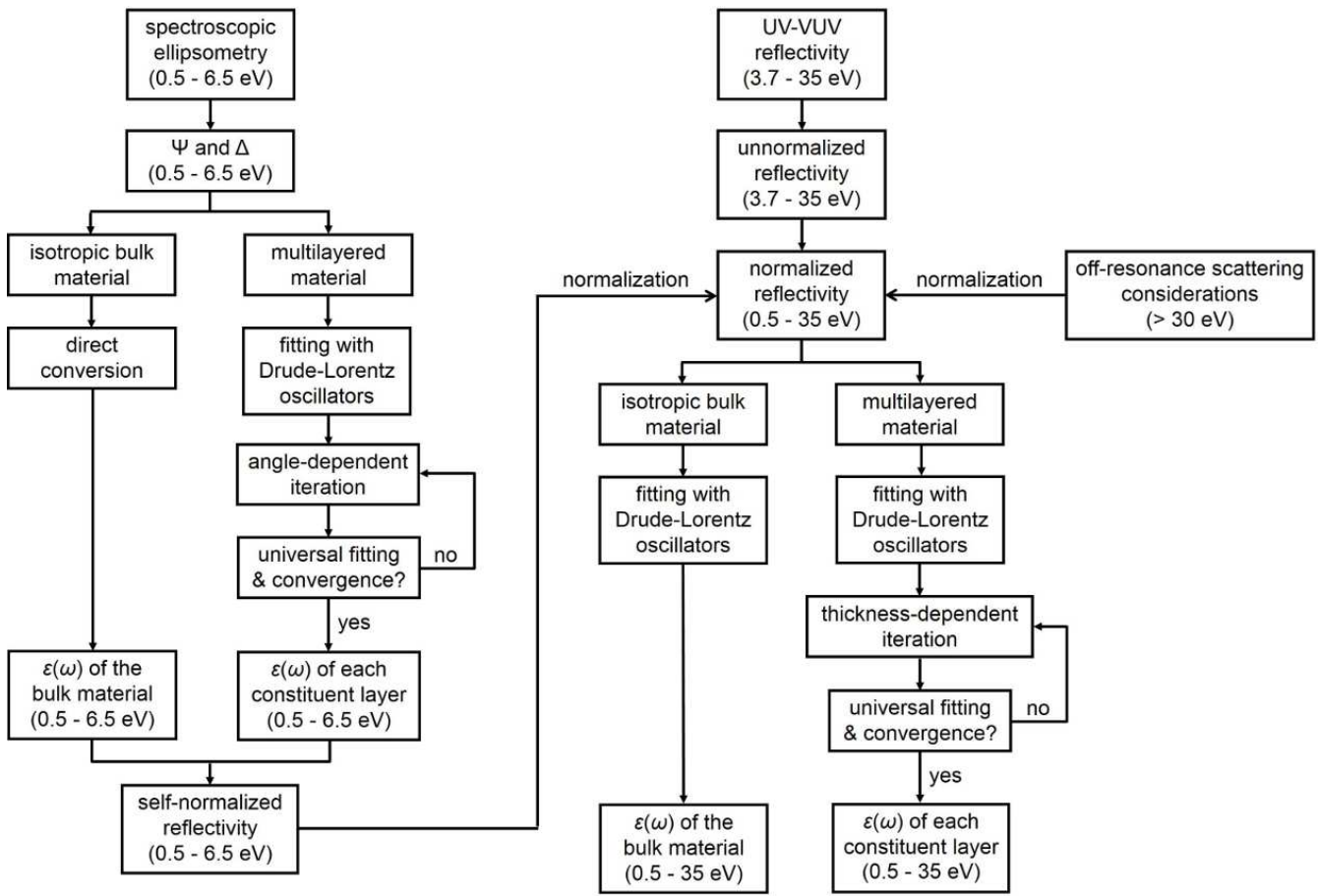}
\caption{Flowchart diagram depicting the whole analysis process of spectroscopic ellipsometry and UV-VUV reflectivity data to obtain the complex dielectric function of both bulk and multilayered materials.}
\label{fig:fig18}
\end{figure*}

In conclusion, we have discussed various aspects of optical analysis of bulk and multilayered materials, as summarized by the flowchart in Figure~\ref{fig:fig18}. For analysis of multilayered materials, we present the self-consistent iteration procedure as a useful tool to separate and extract the dielectric functions of each individual layer. The method is based on the two main variables that affect the reflectivity of a multilayered system: photon incident angle (angle-dependent) and layer thickness (thickness-dependent). By measuring the samples at different incident angles or on samples with slightly different layer thicknesses, self-converged iteration can be performed by cycling through these differently-measured spectra. With enough iteration steps, stabilized separation and extraction of dielectric function of each individual layer can be achieved. By applying the procedure into spectroscopic ellipsometry and UV-VUV reflectivity data of LaAlO$_{3}$/SrTiO$_{3}$, we are able to separate the effects of the interface layer from the LaAlO$_{3}$ film and the SrTiO$_{3}$ substrate. With proper adjustments, this method can be extended to other multilayered material systems with various numbers of layers, making it a very versatile tool in analyzing the optical properties of various multilayered systems.

\begin{acknowledgments}
We acknowledge discussions with Michael R\"{u}bhausen, Ariando, and T. Venkatesan.
This research is supported by the National Research Foundation, Prime Minister's Office, Singapore under its Competitive Research Programme (CRP Awards No. 8-2011-06 and NRF2008NRF-CRP002024), MOE-AcRFTier-2 (MOE2010-T2-2-121), NUS-YIA, and FRC.

\end{acknowledgments}

\end{document}